\documentclass[times,linenumbers]{aastex63}
\nolinenumbers
\usepackage{graphicx}
\usepackage{natbib}
\usepackage{color}
\usepackage{wrapfig}
\usepackage{xspace}
\usepackage{amsmath}
\usepackage{hyperref}  

\usepackage{booktabs}
\usepackage[normalem]{ulem} %DIF > 
\newcommand{\mj}{\ensuremath{\,M_{\rm Jup}}}
\newcommand{\mearth}{$M_\oplus$\xspace}
\newcommand{\mum}{$\mu$m}

\newcommand{\hd}{HD~80606\xspace}
\newcommand{\hdb}{HD~80606~b\xspace}
\newcommand{\hdc}{HD~80606~c\xspace}

\DeclareOldFontCommand{\bf}{\normalfont\bfseries}{\mathbf} %DIF PREAMBLE
\RequirePackage{letltxmacro} %DIF PREAMBLE

\begin{document}
% \submitjournal{AAS Journals}
\shorttitle{HD80606 Orbit timing}
\shortauthors{Beichman et al.}

% \title{Searching for Planets Orbiting Our Closest Solar-Type Neighbor}
\title{Eclipse Timing of the Eccentric Planet \hdb with JWST: Constraints on a Second Planet and other Dynamical Effects}

\correspondingauthor{Charles Beichman}
\email{chas@ipac.caltech.edu}

\author[0000-0002-5627-5471]{Charles Beichman}
\affiliation{NASA Exoplanet Science Institute, IPAC, Pasadena, CA 91125}
\affiliation{Jet Propulsion Laboratory, California Institute of Technology, Pasadena, CA 91109, USA}
\author[0000-0001-8382-7051]{Billy Quarles}
\affiliation{Department of Physics and Astronomy, East Texas A\&M University Commerce, TX 75428, USA}
\author[0000-0003-3504-5316]{B.J. Fulton}
\affiliation{NASA Exoplanet Science Institute, IPAC, Pasadena, CA 91125}

\author[0000-0002-8507-1304]{Nikole Lewis}
\affiliation{Dept. of Astronomy, Cornell University, Ithaca, NY}

\author[0000-0002-8211-6538]{Ryan  Challener}
\affiliation{Dept. of Astronomy, Cornell University, Ithaca, NY}

\author[0000-0003-2415-2191]{Julien de Wit}
\affiliation{Dept. of Earth, Atmospheric and Planetary Science, Massachusetts Institute of Technology: Cambridge, MA, US }

\author[0000-0002-7670-670X]{Malena Rice}
\affiliation{Department of Astronomy, Yale University, 219 Prospect Street, New Haven, CT 06511, USA}

\author[0000-0001-9164-7966]{Julie Inglis}
\affiliation{ Department of Astronomy \& Astrophysics, University of California San Diego, 9500 Gilman Dr. La Jolla, CA 92093}

\author[0000-0003-3759-9080]{Tiffany Kataria}
\affiliation{Jet Propulsion Laboratory, California Institute of Technology, Pasadena, CA 91109, USA}

\begin{abstract}

A variety of effects can perturb the orbital properties of  single planets in close orbits around their host stars. HD80606 b is a highly eccentric  ($\epsilon$=0.93) exoplanet  orbiting  its host G5V star, \hd. \hdb became a compelling target for exoplanet research after \citet{Laughlin2009} discovered that the 4.16 \mj\ planet transits and is eclipsed by \hd, thereby enabling photometric and spectroscopic observations to yield a wealth of information about the object. The exquisite precision of recent  JWST eclipse timing offers an opportunity to investigate whether the orbit of \hdb is modified due to a variety of mechanisms, including  General Relativity, tidal torques, or the presence of a second perturbing planet. We have used over 25 years of radial velocity  data plus eclipse and transit observations to place limits on the precession of \hdb's orbit and to assess which of these effects, if any, are observable.  The new models are consistent with constant values of period, eccentricity and the argument of periastron, $\omega$, with a limit on $\dot{\omega}$ at the level of the predicted GR drift. The PRV and timing data limit the mass and location of a second planet external to \hdb. 

Timing offsets  between JWST eclipses relative to the \citet{Pearson2022} predictions are attributed to the poor constraints on $\sqrt{\epsilon} (\cos\ \omega, \sin\ \omega)$ with only the single eclipse measurement available (Spitzer 2009)  in that analysis  and to an over-weighting of the eclipse timing in that earlier analysis.

\end{abstract}

\section{Introduction}

\begin{figure}[!t]
\centering
\includegraphics[width=0.6\textwidth]{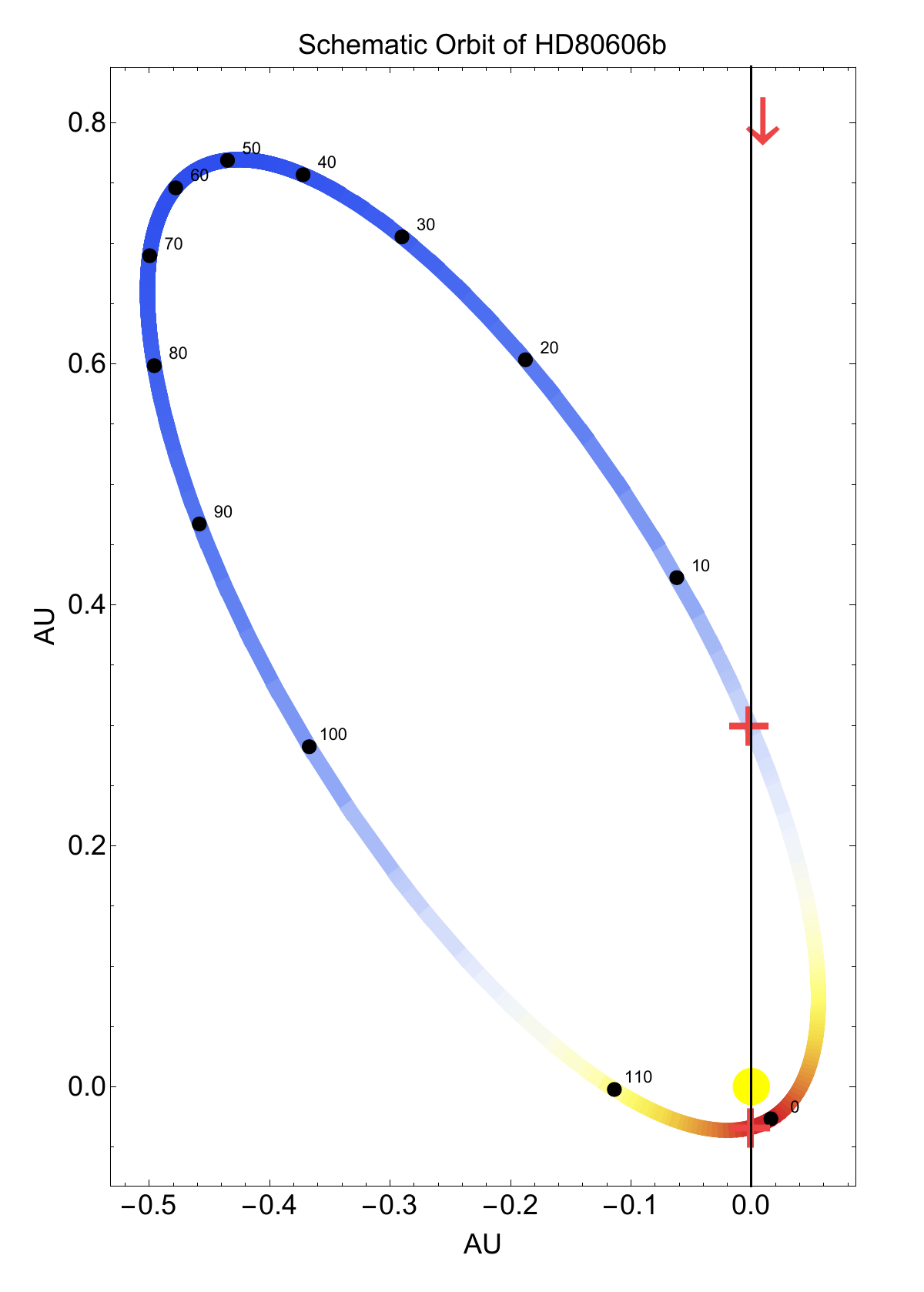}
\caption{A schematic of the highly eccentric orbit of \hdb. The arrow indicates the direction of observation with '+' marks denoting the midpoints of the transit and secondary eclipse. Tick marks are given at 10 day intervals with the color of the line giving a rough indication of temperature, $T\propto r^{-0.5}$, from 400 K to over 1400 K. \label{fig:orbit}}
\end{figure}

 A variety of effects can perturb the orbits of even single planets, particularly those in orbits close to the host star. Some of these effects include the General Relativistic (GR) induced precession of the orbit and tidal effects during periapsis passage as summarized by \citet{Pal2008, Jordan2008, Blanchet2019}.   For example, the  decaying orbit of WASP-12b (P= 1.09 d)   has been attributed to tidal interactions \citep[and refs therein]{Yee2020}, while WASP-4b (P=6.1 d) shows marginal evidence for orbital decay or apsidal precession \citep{Turner2022}. Such systems hold the promise of determining the internal properties of  the planet if orbital effects can be attributed to the effects of tidal dissipation.

One planet that has been suggested as a strong candidate for potential orbital changes over observational baseline timescales is HD 80606 b. The 4.6 \mj\ planet \hdb (P=111.4 d, e=0.932), orbits the G5V star HD 80606 and is the second most eccentric transiting exoplanet after the recently discovered TIC 241249530 b (P=165.8 d, e=0.941; \citet{Gupta2024}). HD 80606 b has a long history of transit, eclipse and RV observations extending back almost two decades to its discovery as an eclipsing system by \citet{Laughlin2009}. The properties of \hd and \hdb are given in Table~\ref{tab:stellar}. Figure~\ref{fig:orbit} presents a schematic of \hdb's 111.4d orbit showing the alignment that produces a long transit (12 hr) and a much shorter eclipse (1.7 hr).

\hdb's high eccentricity  brings it close enough to its host star for tidal, GR and other effects to become strong enough to produce  observable effects.  Recently obtained JWST eclipse timing data is 10-20 times more precise than the earlier transit and eclipse observations, thus permitting a much improved characterization of \hdb's orbit. Combining these new constraints with archival datasets we use measurements of \hdb's orbit, including more than 25 years of Precision Radial Velocity (PRV) observations  plus  transit  and eclipse timing data, to search for small orbital perturbations and the mechanisms which might cause them.

In $\S$2 we summarize the available PRV,  eclipse and transit data and introduce one new transit observation from the CHEOPS mission and two eclipse observations from JWST. 
In $\S$\ref{sec:models} we fit an orbital model  to the data and find that a fit with constant parameters ($T_0,\, e, \, P,\, \omega$) adequately represents the data. A model with a linear changing $\omega$ is consistent with the data but is not required. The differences seen with respect to the \citet{Pearson2022} results are attributed to the limited available eclipse data and the particular assumptions of that modeling. In $\S$\ref{sec:effects} we briefly discuss the magnitude of various effects which might change the orbit of \hdb. In $\S$\ref{sec:second_planet} we set limits on the mass and orbital location of any perturbing planet and discuss the prospects for future observations.  $\S$\ref{sec:conclusion} presents  concluding remarks.

\begin{deluxetable}{llr}
\centering
\tabletypesize{\scriptsize}
\tablecaption{System Parameters for HD~80606\label{tab:stellar}}
\tablehead{
\colhead{Parameter} & \colhead{Explanation, reference} & \colhead{Value} }
\startdata
\multicolumn{3}{l}{\textit{HD 80606}}\\
M$_*$ [M$_\odot$] & Stellar Mass (1) & 1.047$\pm$0.047 \\
R$_*$ [R$_\odot$] & Stellar Radius (2) & 1.050 $\pm$ 0.01 \\
T$_*$ [K] & Stellar Temperature (3) & 5584 $\pm$ 13\\ 
G & Gaia mag &8.83\\ 
Dist & distance (pc) & 66.0\\\hline
\multicolumn{3}{l}{\textit{HD 80607}}\\
M$_*$ [M$_\odot$] & Stellar Mass & 1.0 \\
T$_*$ [K] & Stellar Temperature (3) & 5506 $\pm$ 15\\ 
G (mag)& Gaia mag &8.96\\ 
Sepn & Projected HD 80606/7 Separation (AU) & 1360\\ \hline
\multicolumn{3}{l}{\textit{Illustrative values for HD 80606b}}\\
$R_p$ [R$_{\rm Jup}$] & Planet Radius (2) & 1.032$\pm$0.015 \\
$M_{p}$ [M$_{\rm Jup}$] & Planet Mass (2) & 4.164$\pm$ 0.005 \\
K [m/s] & RV Semi-Amplitude (2) & 469.22 $\pm$ 0.61 \\
Period [day] & Orbital period (2) & 111.436 \\
$i$ [deg] & Inclination (2) & 89.24 \\
a [au] & Semi-major axis (2) & 0.460 \\
$\epsilon$ & Eccentricity (2) & 0.932\\
\enddata
\tablecomments{Refs: 1, \citet{Rosenthal2021}; 2, Illustrative values from \citet{Pearson2022}. More precise values are given below based on fitting the new data; 3) \citet{Liu2018} state that HD80607 is only slightly less massive than HD 80606. For our purposes we take the mass of HD 80607 to be 1 M$_\odot$.}
\end{deluxetable}

\section{Transit and Eclipse Observations of HD80606~b }

Table~\ref{tab:timing_data} incorporates the transit and eclipse observations described in \citet{Pearson2022} to which we have added: 1) a re-reduction of the \citet{Laughlin2009}  eclipse observed by Spitzer at 8.0 \mum\ in 2007 and a second eclipse observed in 2010 \citep{dewit2016} at 4.5 \mum; 2) a transit observed by the CHEOPS mission in 2021; 3)  the two eclipses  observed with JWST the first with  NIRSpec in 2022 \citep{Sikora2025}  and   the second by MIRI-LRS  in 2023 (Kataria et al, in prep.). The two Spitzer measurements were discussed in \citet{dewit2016} but no timing information was provided there. This is discussed in $\S$\ref{sec:newSpitzer}.

The CHEOPS transit measurements have not been presented elsewhere and we describe a fit to those data below ($\S$\ref{sec:CHEOPS}). We adopt the timing information for JWST/NIRSpec from
\citet{Sikora2025}. We analyze the  timing data for JWST/MIRI-LRS spectroscopy (Kataria et al, in prep), leaving the detailed description of those observations and their spectral content to a separate paper. The 1$\sigma$ precision of the JWST measurements is remarkable, 42 sec for NIRSpec and 12 sec for MIRI which can be compared with a 200-300 sec for the earlier transit  values. The two  Spitzer eclipse measurements offer excellent precision although limited by Spitzer's smaller collecting area compared with JWST.

The times in the Table~\ref{tab:timing_data} are presented in BJD and are thus corrected for location of the Earth in its orbit. The Spitzer eclipse time is given in HJD, with a small $\approx$4 sec correction  made to turn into BJD. However, for the purposes of comparing transit and eclipse timing, these times must be adjusted to a \hd reference frame by accounting for the light travel time between \hdb and Earth. Examination of Figure~\ref{fig:orbit} shows that the transit occurs when \hdb  is in front of the star and closer to Earth, whereas  the eclipse occurs with  \hdb   behind the star and further from Earth.  As discussed below, the corrections are typically less than 1/3$\sim$1/2 of the timing uncertainties, except for the JWST/MIRI observation for which  $\Delta T_{corr}$ is comparable to the timing uncertainty.

\begin{table}[h]
\centering
\caption{Transit and Eclipse Timing Data}
\label{tab:timing_data}
\begin{tabular}{lrcccc}
\toprule
Type & Orbit & BJD & Unc.\ (d) & Facility & {\edit1{Central Wavelength}} \\
\midrule
%Eclipse  & $-40$ & 2454424.736    & 0.003\phantom{0} & Spitzer$^2$ \\
Eclipse  & $-40$ & 2454424.7371   & 0.0013\phantom{0} & Spitzer Ch2$^2$ & 4.5 $\mu$m \\
Transit  & $-35$ & 2454987.78424  & 0.0049 & Ground$^3$ &Various visible\\
Eclipse  & $-33$ & 2455204.79763  &0.0009 & Spitzer Ch1$^4$ & 3.6 $\mu$m \\
Transit  & $-33$ & 2455210.6500   & 0.0064 & Ground$^5$ &Various visible\\
Transit  &   $0$ & 2458888.07466  & 0.0020 & TESS$^6$ & 786.5 nm \\
Transit  &   $3$ & 2459222.38613  & 0.0041 & CHEOPS$^7$ & 715 nm \\
Transit  &   $6$ & 2459556.70070  & 0.0035 & Ground$^8$ &Various visible\\
Eclipse  &   $9$ & 2459885.16675  & 0.0005 & JWST/NIRSpec$^9$&2.9 to 5.2 $\mu$m \\
Eclipse  &  $17$ & 2460776.66341  & 0.00015 & JWST/MIRI$^{10}$&5.1 to 11.9 $\mu$m \\
\bottomrule
\end{tabular}
\begin{flushleft}
\small
\textbf{Notes:}
We have taken timing values only for full transits for this analysis. $^1$Orbit numbers given with respect to the time of periapsis for Orbit \#0  \citep{Pearson2022}; $^2$The 2007 8.0 \mum observation obtained by \citet{Laughlin2009,dewit2016} and re-reduced as described in $\S$\ref{sec:newSpitzer}; $^3$\citet{Winn2009};$^4$A 4.5 \mum (Ch1) observation obtained by \citet{dewit2016} and reduced as described in $\S$\ref{sec:newSpitzer};
$^5$\citet{Shporer2010}; $^6$\citet{Pearson2022};$^7$CHEOPS; see $\S$~\ref{sec:CHEOPS}; $^8$\citet{Pearson2022}; $^9$JWST/NIRSpec \citep{Sikora2025}; $^{10}$ JWST/MIRI-MRS (Kataria et al, in prep). As discussed in the text, these values are in the BJD heliocentric reference frame, but have not been corrected not for motions within \hd. Light travel time corrections are discussed in the text.
R{\o}mer delay corrections ($+150$\,s for transits, $-17$\,s
for eclipses) are applied in the model (Section~1.4).
\end{flushleft}
\end{table}

\begin{figure*}[!t]
\centering
\includegraphics[width=0.8\textwidth]{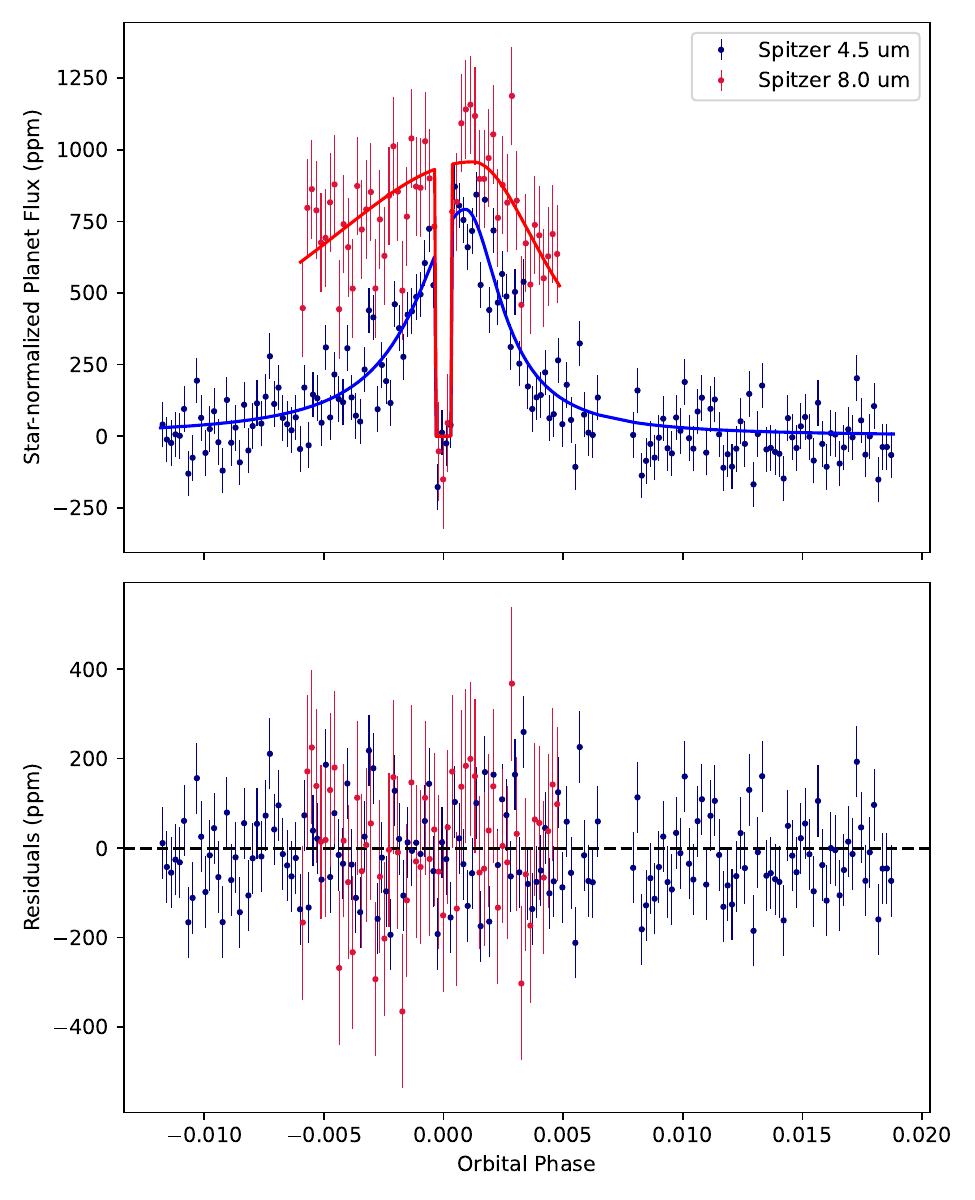}
\caption{Model fits to the 2007 and 2010 eclipses at 8.0 and 4.5 \mum\ respectively.\label{fig:Spitzer}}
\end{figure*}

\begin{figure*}[!t]
\centering
\includegraphics[width=0.45\textwidth]{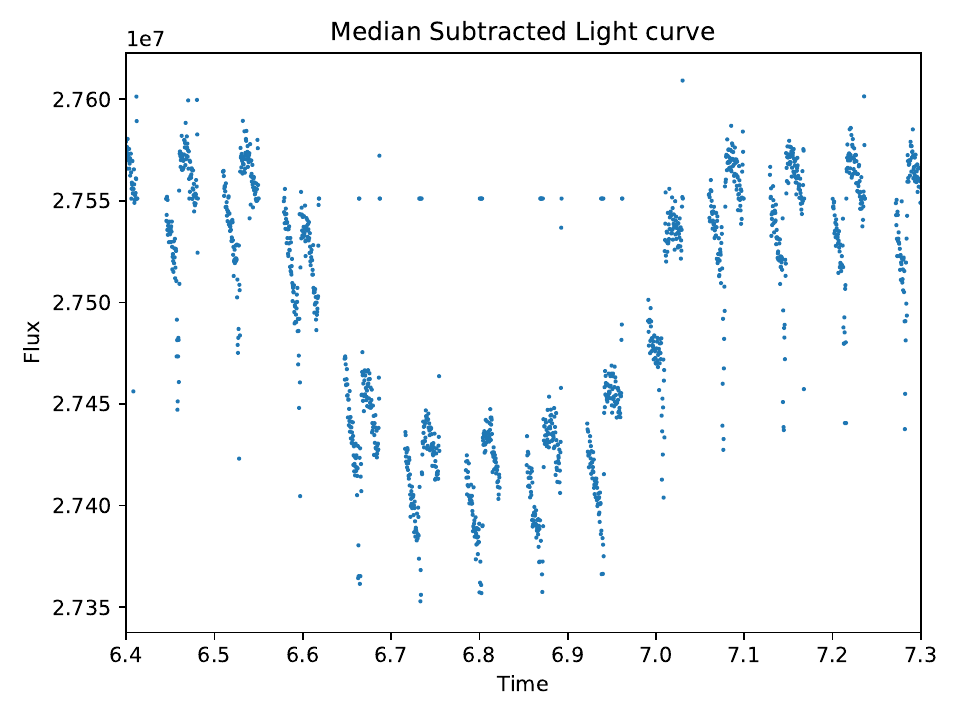}\includegraphics[width=0.45\textwidth]{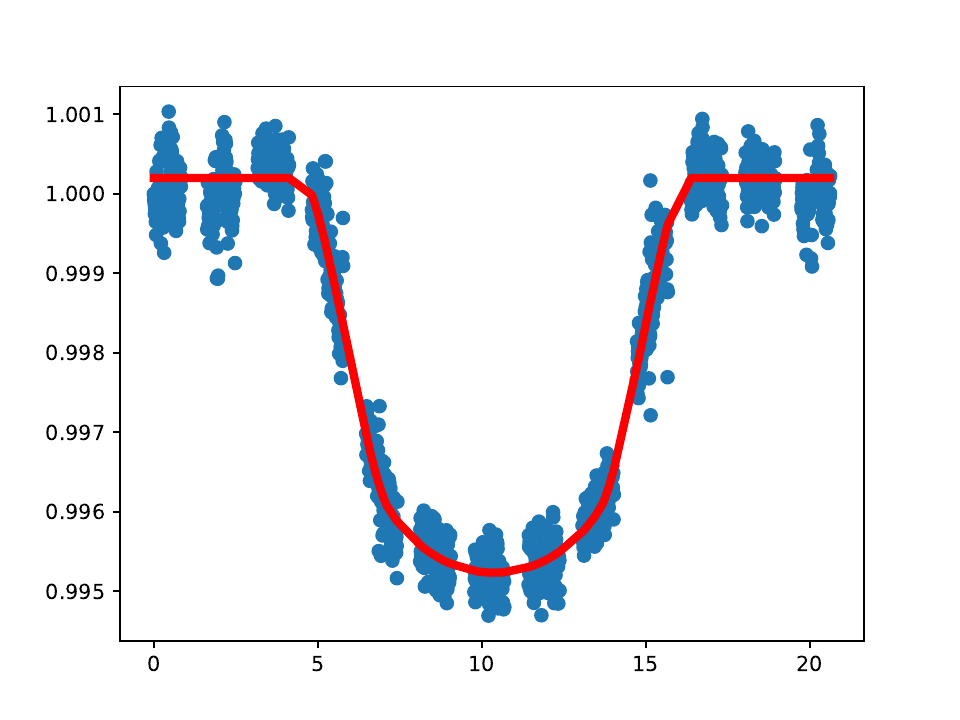}
\caption{left) The Level-2 light curve obtained from CHEOPS archive for a transit occurring on 2021-1-08. right) Cleaned light curve corrected for outliers and the hook- shaped artifact. The red curve denotes the EXOFAST fitted model used to derive the timing of the eclipse.\label{fig:CHEOPS}}
\end{figure*}

\begin{figure*}[!t]
\centering
\includegraphics[width=0.9\textwidth]{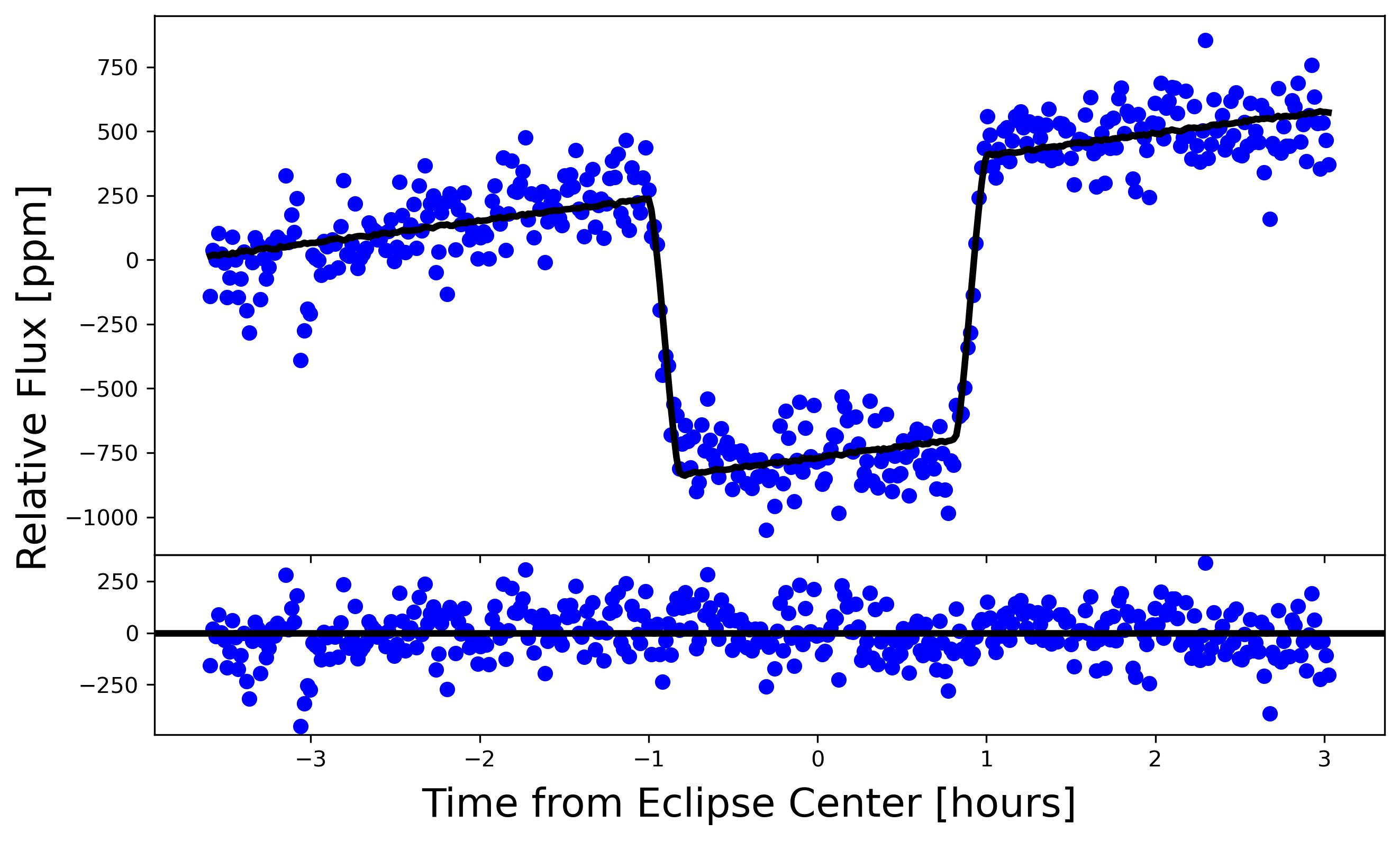}
\caption{MIRI LRS 8~$\mu$m eclipse light-curve with astrophysical and systematics fit (top) and residuals to that fit (bottom). The observations have been binned to a 1 minute cadence for clarity. The linear trend with time across the eclipse is due to variations in time of the flux of the planet as it heads toward periastron passage. \label{fig:MIRI}}
\end{figure*}

\subsection{Spitzer Eclipse Measurements \label{sec:newSpitzer}}

Spitzer made two eclipse observations of \hdb,  the original detection by \citet{Laughlin2009} in Ch2 (8.0 \mum) and one in 2010 in Ch1 (4.5 \mum). We present  a re-analysis of the timing of the 2007 measurement and the first analysis of the 2010 observation timing.

The Spitzer/IRAC 4.5 and 8.0 \mum\ photometry of HD 80606 b presented respectively in \citet{dewit2016,Laughlin2009}  was obtained from extensive time-series observations around periastron passage and reduced using a standard aperture photometry pipeline including background subtraction, centroid determination, and optimization of aperture size to minimize both white and red noise. The resulting light curves were then corrected for the dominant instrumental systematics (most notably intrapixel sensitivity variations and detector ramp effects) within a global MCMC framework in which these systematics were fitted jointly with the planetary signal following the method introduced in \citet{dewit2012}. This approach yields systematics-corrected light curves (shown in Figure 2 of \citet{dewit2016}) that reach close to the photon-noise limited precision. Importantly, the instrumental effects as well as the planet’s transient heating operate on timescales distinct from the sharp ingress/egress of the eclipse, such that the timing information is largely preserved in the corrected data product. In the original analysis, the data were modeled in orbital phase space within a global atmospheric framework, with the orbital configuration held fixed, and no explicit fit was performed for individual eclipse mid-times. As a result, the eclipse timings were not directly extracted, motivating a dedicated re-fit of the corrected light curves (as used here) with the eclipse times allowed to float.

\subsection{CHEOPS Transit \label{sec:CHEOPS}}
The CHEOPS spacecraft \citep{Benz2021} observed a full transit of \hdb on orbit \#3 in our nomenclature (2021-01-01)\footnote{'CH\_PR110044\_TG001001\_TU2021-01-04T21-58-02\_SCI\_COR\_Lightcurve-R40\_V0300.fits'} as shown in Figure~\ref{fig:CHEOPS}a. The 52 Level 2 (L2) data segments obtained during a 4-day observing sequence from the CHEOPS archive were first corrected for outliers using n-sigma rejection. A template for the hook-shaped artifact which occurs on each CHEOPS orbit during target acquisition was derived from the out-of-eclipse data and subtracted from all the observations leading to the clean light-curve shown in Figure~\ref{fig:CHEOPS}b. The EXOFAST 2 software \citep{Eastman2013} available at the NASA Exoplanet Archive was used incorporating priors from \citet{Pearson2022} and limb-darkening coefficients derived from the Exoplanet Characterization Toolkit\footnote{https://exoctk.stsci.edu/limb\_darkening} for the CHEOPS passband. The fit also incorporated radial velocity data from the HIRES instrument as described in \citet{Pearson2022}. The goal of the fitting was not to re-derive properties of the planet, which are already well-determined by the analysis of multiple events \citep{Pearson2022}, but rather to compute the center-time of the transit as accurately as possible (Table~\ref{tab:timing_data}).

\subsection{JWST MIRI Eclipse}
We used the \textsc{Eureka!} data reduction pipeline \citep{Bell2022} to determine the eclipse timing from observations of HD~80606b's partial phase curve observations with JWST MIRI LRS (JWST-GO-2008, PI Kataria). The \textsc{Eureka!} data reduction pipeline was specifically designed for exoplanet time-series observations and has been used in over 100 JWST observational studies of exoplanets and their atmospheres. We begin our \textsc{Eureka!} reduction of the observations from the {\it uncal.fits} files and leverage the standard Stage 1 and 2 JWST MIRI time-series observation data reduction settings. In Stage 3 we perform background subtraction and spectral extraction on the individual integration frames ({\it calints.fits} with a background exclusion region 10 pixels and an aperture half-width of 4 pixels from the central pixel of the trace. In Stage 4, we extracted a single spectroscopic channel spanning 6.408 to 9.338~$\mu$m to replicate the wavelength coverage of {\it Spitzer} IRAC Ch4. Finally, in Stage 5 we isolate the eclipse portion of our partial phase-curve by clipping between integrations 10,000 and 20,0000 and then perform a fit to the eclipse using a combination of the \textsc{batman} \citep{Kreidberg2015} and \textsc{emcee} \citep{FM2013}. In fitting the astrophysical (planetary eclipse) signal, we hold the orbital period ($P$), time of transit ($T_0$), orbital inclination ($i$), orbital distance ($a/R_{\star}$), orbital eccentricity ($e$) and argument of periastron ($\omega$) fixed to the values from \citet{Pearson2022} and allow the time and depth of secondary eclipse to float. We include a linear with time trend to account for the planetary phase variation across the eclipse\footnote{Note that since the eclipse occurs several hours after the start of the observations, it is unnecessary to account for the exponential ramp that commonly occurs at the beginning of MIRI LRS bright-object time-series observations.}, account for correlations between the flux and trace position on the detection, and include an error inflation term.

Figure~\ref{fig:MIRI} shows our fit to our JWST MIRI 8~$\mu$m eclipse of HD~80606b. We derive a secondary eclipse time of 60776.16241$\pm$0.00015 MBJD$\_$TDB and an eclipse depth of 1090$\pm$12~ppm. Our derived MIRI eclipse time differs by 28 minutes from the value in Table~8 \citep{Pearson2022} and a similar temporal offset was noted in the JWST NIRSpec G395H eclipse by
\citet{Sikora2025}. The origin of this temporal discrepancy is discussed in $\S$\ref{sec:fitting}. Our eclipse depth is larger and more precise, but within $3\sigma$, of the HD~80606b 8.0~$\mu$m eclipse depths reported in \citet{dewit2016} (1053$\pm$94~ppm) and \citet{Laughlin2009} (1000$\pm$200~ppm). In Kataria et al., (in prep), we will use  the full spectroscopic capabilities of the MIRI/LRS to derive an emission spectrum of HD~80606b and investigate variations in eclipse timings with wavelength. 

In this paper, we present a relatively simple secondary eclipse fit to the MIRI LRS observations of HD 80606 using a linear baseline combined with the secondary eclipse model. As a result of the linear baseline, the base of the secondary eclipse is not flat in this model. We have compared this fit to a more complicated asymmetric Lorentzian model of the entire phase curve plus the secondary eclipse adopted for a subsequent paper (Kataria et al. submitted). We additionally tested allowing the orbital parameters, $a/R_*$, period, inc, eccentricity and omega to vary during the fit rather than fixing these parameters to previously derived values. We find that the secondary eclipse midpoint does not vary by more than 1-2 sigma in any of these variations, nor is the precision of the derived time affected.  The simplifying assumptions adopted  here do not affect the upper limits  derived herein.

\subsection{Radial Velocity Data}

We have obtained 187 radial velocity measurements of HD~80606 from three
instrument configurations (Table~\ref{tab:PRV_data}): APF (72), Keck/HIRES post-upgrade (76),
and Keck/HIRES pre-2004 (39), excluding all KPF data.
Each instrument is assigned an independent zero-point ($\gamma$)
and jitter term ($\sigma_\mathrm{jit}$) added in quadrature to
the formal uncertainties.

\begin{deluxetable}{lccc}
\centering
\caption{Radial Velocity Data}
\label{tab:PRV_data}
\tablehead{
\colhead{BJD} & \colhead{RV (m s$^{-1}$)} & \colhead{$\sigma$(RV) (m s$^{-1}$)	}&\colhead{Instrument} }
\startdata
2452007.898	&	-38.108275	&	2.055665	&	HIRES\_pre2004	\\
2452219.162	&	-4.881094	&	1.679194	&	HIRES\_pre2004	\\
2452236.059	&	-56.638495	&	1.850351	&	HIRES\_pre2004	\\
2452243.168	&	-75.863486	&	1.719186	&	HIRES\_pre2004	\\
2452307.879	&	611.831552	&	1.821998	&	HIRES\_pre2004	\\
2452333.01	&	-8.259299	&	1.972414	&	HIRES\_pre2004	\\
2452334.015	&	-20.199118	&	1.784482	&	HIRES\_pre2004	\\
2452334.905	&	-20.48207	&	1.594249	&	HIRES\_pre2004	\\
\enddata
\tablecomments{
Complete table is available online: \dataset[https://doi.org/10.5281/zenodo.20096135]{https://doi.org/10.5281/zenodo.20096135
}}
\end{deluxetable}

\section{Model Fitting \label{sec:models}}

The orientation of \hdb's orbit as seen from Earth (Figure~\ref{fig:orbit}) means that the eclipse precedes periastron passage with the transit following   some days later. Equation~\ref{eqn:timing} describes the offset between time of periastron passage, $T_{Peri}$ and the times of the midpoints of the transit $T_{tr}$ and eclipse $T_{ecl}$ \citep{Huber2017,Pearson2022}: 

\begin{equation}
\Delta T = \frac{P}{2 \pi \sqrt{1-\epsilon^2}}\ \int_{0}^{\nu}\left (\frac{1-\epsilon^2}{\epsilon \cos (x)+1}\right)^2 dx \label{eqn:timing}
\end{equation}

\noindent where $P$ is the orbital Period, $\epsilon$ is the eccentricity, and the true anomaly, $\nu$, is given by $\pi/2-\omega$ (Transit) and $-\pi/2-\omega$ (Eclipse), and $\omega$ is the argument of periastron. For \hdb the typical value for $T_{Transit}-T_{Peri}=\Delta T_{Transit}$ is 5.73 days and $T_{Eclipse}-T_{Peri}=\Delta T_{Eclipse}$ is -3.1 hours, i.e. with the eclipse preceding Periapsis. This equation is used in conjunction with a full orbital model to predict the Doppler reflex motion for comparison with 15 years of Lick/APF and Keck/ HIRES PRV measurements \citep{Fulton2018,Rosenthal2021,Pearson2022}. 

\subsection{Orbital Model}

To evaluate the model parameters we use 187 radial velocity measurements,  9 mid-transit and mid-eclipse times spanning orbits
$-40$ to $+17$ ($\sim$17 years; Table~\ref{tab:timing_data}):
5 transits (ground-based, TESS, CHEOPS) and 4 eclipses (2 Spitzer,
JWST/NIRSpec, JWST/MIRI).
The most precise is the JWST/MIRI eclipse at orbit~17
($\sigma = 13$\,s).
We fit the RV and timing data simultaneously with a Keplerian
orbit parameterized by $T_c$, $P$, $e$, $\omega_0$, and $K$,
using the orbital conventions and RV model from
\textsc{RadVel} (Fulton et~al.\ 2018).
We fit directly in the physical parameters rather than
the $\sqrt{e}\cos\omega$, $\sqrt{e}\sin\omega$ basis,
as the high eccentricity ($e = 0.93$) ensures well-behaved
posteriors in $e$ and $\omega$.
The argument of periastron $\omega_0$ is defined at
$T_\mathrm{ref} = 2458882.344$\,BJD.
Transit and eclipse times follow from Kepler's equation
with $f_\mathrm{tra} = \pi/2 - \omega$ and
$f_\mathrm{ecl} = -\pi/2 - \omega$.
In the precessing model (Case~D), $\omega$ evolves linearly:
$\omega(t) = \omega_0 + \dot\omega\,(t - T_\mathrm{ref})$.

\subsection{Light Travel Time Correction}

We apply a R{\o}mer delay correction for the finite light travel
time across HD~80606~b's orbit.
The line-of-sight displacement is
$\Delta z = r\sin i\,\sin(f + \omega)$,
where $r = a(1-e^2)/(1+e\cos f)$ and $i = 89.269^\circ$.
The predicted observed time is
$T_\mathrm{obs} = T_\mathrm{bary} - \Delta z / c$.
At transit (near apastron) the correction is $\sim$150\,s;
at eclipse (near periastron) it is $\sim$17\,s.
The JWST/MIRI eclipse correction exceeds its measurement
uncertainty and must be included.

\subsection{Fitting Procedure \label{sec:fitting}}

We consider four cases:
A (RV only, $\dot\omega = 0$),
B (RV + 7 pre-JWST timing, $\dot\omega = 0$),
C (RV + all 9 timing, $\dot\omega = 0$), and
D (same as C with $\dot\omega$ free).
We find MAP solutions via Levenberg--Marquardt optimization,
then sample posteriors with   \href{https://github.com/dfm/emcee}{\texttt{emcee}}
(Foreman-Mackey et~al.\ 2013; 50 walkers, 20{,}000 steps,
5{,}000 burn-in, thinned by 10).
We impose a weak Gaussian prior on $\dot\omega$
($\sigma = 5\times10^{-6}$\,rad\,day$^{-1}$;
$|\dot\omega| < 3\times10^{-5}$\,rad\,day$^{-1}$)
and compare models via BIC and AICc.

Details of each parameter fitting are presented in Table~\ref{tab:FitResults}:
\begin{itemize}
 \item Case A: The fits to the RV data only yield only poor constraints on the overall timing of the orbit ($T_{peri}$).
 \item Case B: Fit the PRV data plus the eclipse and transit timing data prior to JWST. This case is equivalent to the model presented in \citet{Pearson2022} and provides a good extrapolation to the JWST epoch with $\chi^2\sim 0.44$ for the timing data and $\chi^2 =1.02$ overall. 
 \item Case C: This model incorporates the PRV data, the seven earlier eclipse and transit measurements, and the two JWST eclipses, yielding a $\chi^2\sim 1.78$ for the 8 timing values and an overall $\chi^2=1.07$.
 \item Case D: This model uses the PRV data, the seven earlier transit and transit measurements, the two JWST eclipses and incorporates a linearly variable $\omega(t)=\omega_0+\dot{\omega}t$. This model yields  a slightly improved fit to the timing data with $\chi^2\sim 1.49$ for the 8 timing points and a net $\chi^2=1.06$ at the cost of an extra free parameter. The value of $\dot{\omega}= 302^{+294}_{-289}$ arcsec century$^{-1}$  is indistinguishable from zero (Figure~\ref{fig:omegadot}). For comparison, the General Relativity prediction (Equation~\ref{eqn:GR}) is $\dot{\omega}_\mathrm{GR} = 203$\,arcsec\,century$^{-1}$.

\end{itemize}

We do not include TTVs from a possible second planet in our orbital fits, but we model them directly with N-body simulations. As discussed below ($\S$~\ref{sec:effects}), a second companion would have to sit beyond 1.5-2 AU, where the maximum TTV values are a few minutes or less. Shifts at this level add scatter about our timing model rather than a secular trend and are small compared to the uncertainty on the fitted precession rate. They also would not affect the RV-based limits on a second planet shown as will be discussed below ($\S$~\ref{sec:second_planet}), and the barycentric light travel time shift is negligible (much less than 1 s).

\begin{figure*}[t]
\centering
\includegraphics[width=0.4\textwidth]{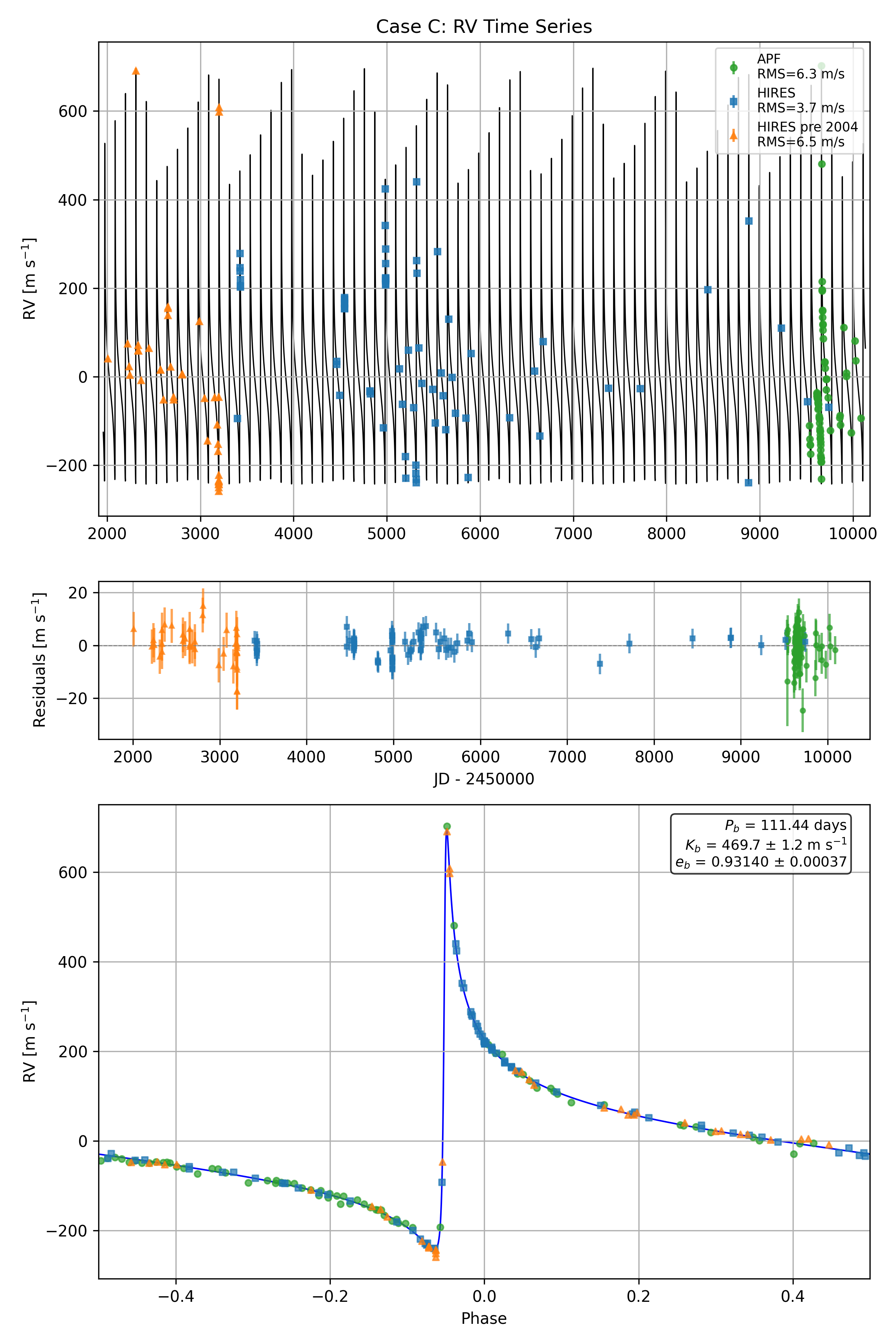}
\includegraphics[width=0.4\textwidth]{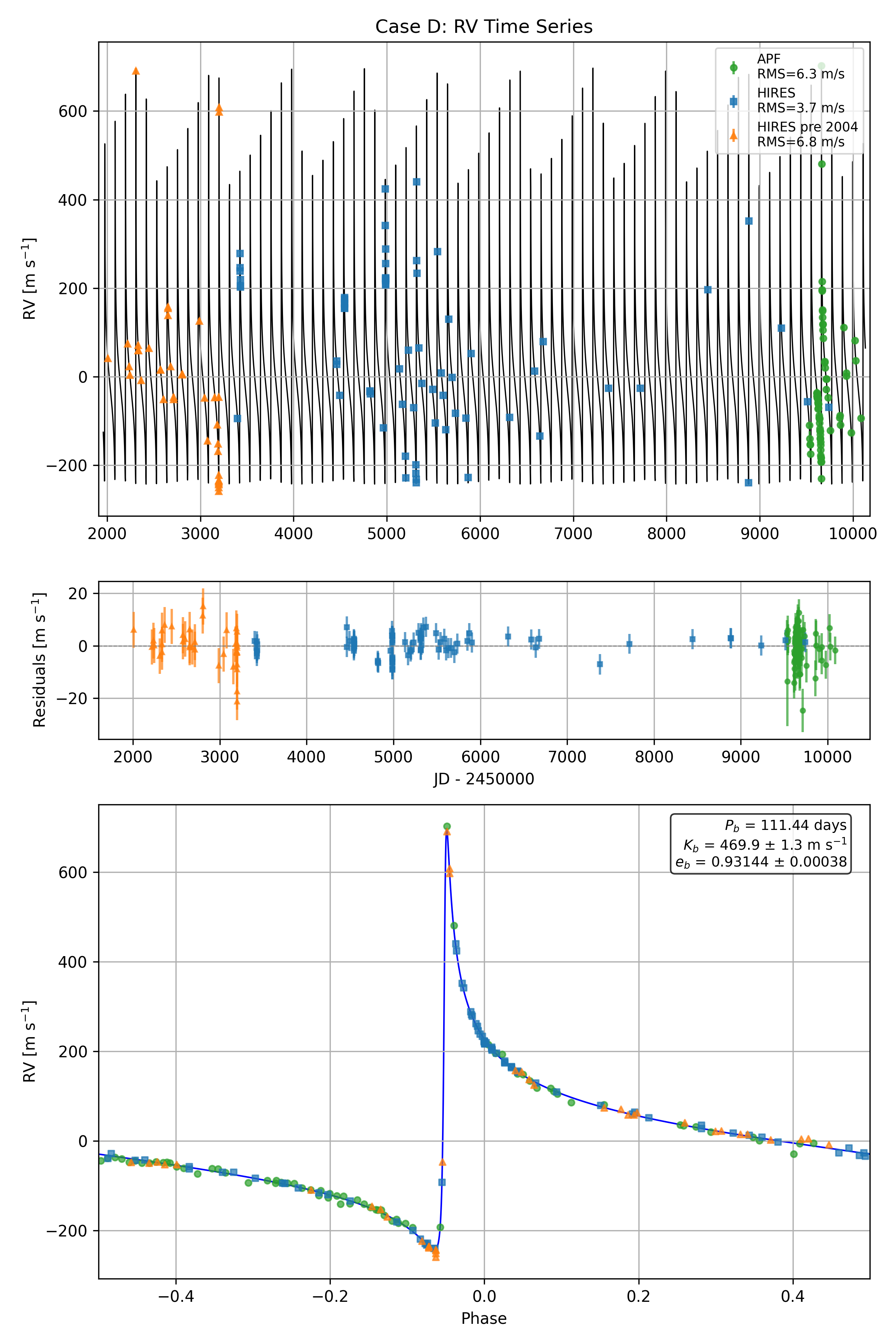}
\caption{Radial velocity orbit fits for Cases C and D.
Each panel shows the time series (top), residuals (middle),
and phase-folded orbit (bottom).
Error bars include instrument jitter added in quadrature.}
\label{fig:rv_orbits}
\end{figure*}

\begin{figure*}[p]
\centering
\includegraphics[width=\textwidth]{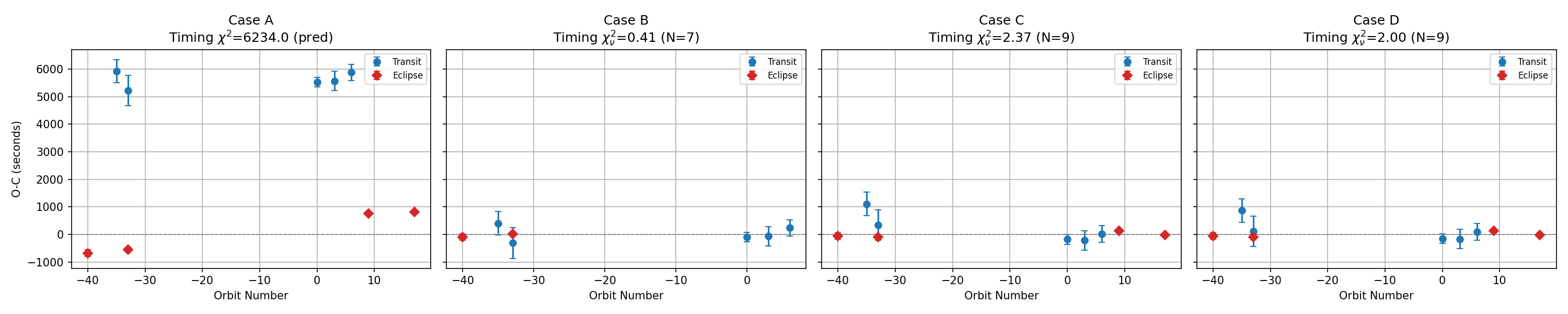}\\
\includegraphics[width=\textwidth]{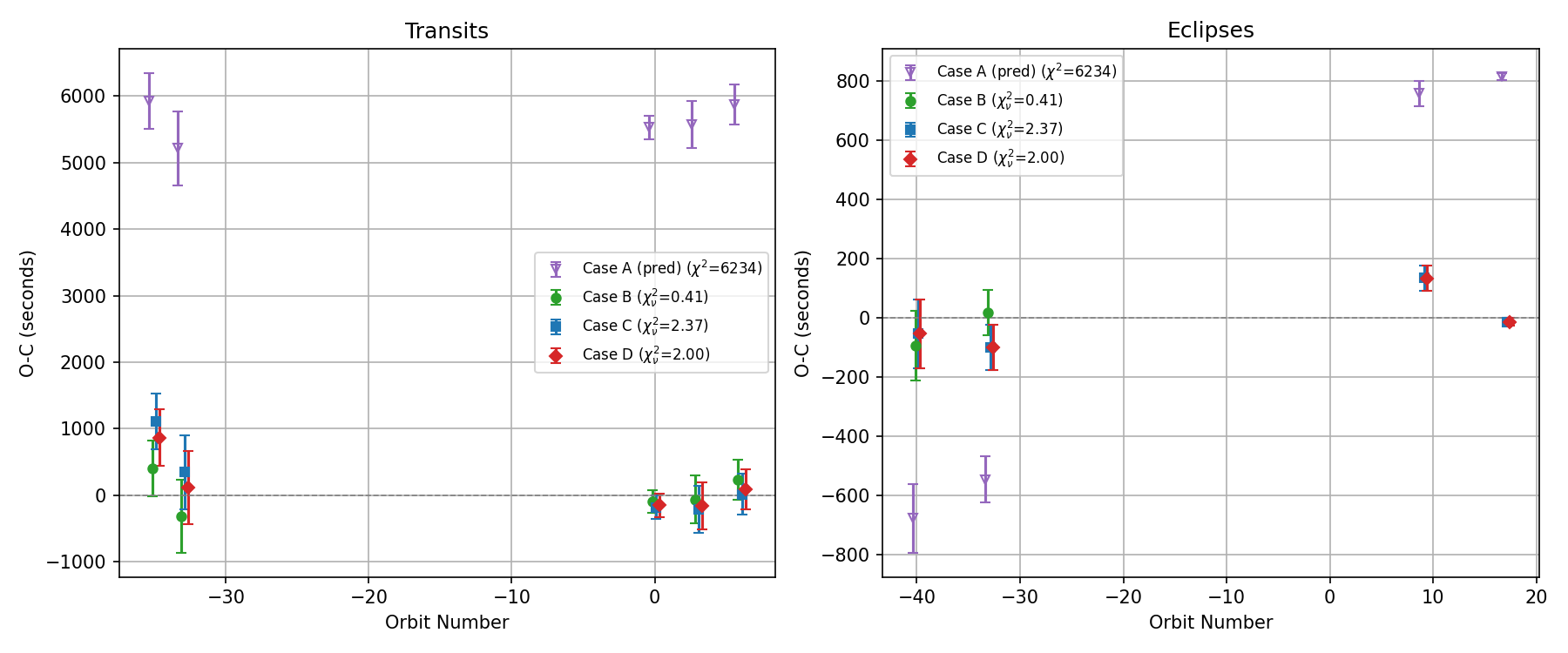}
\caption{top) Observed minus computed (O$-$C) timing residuals for
all four cases at the MCMC median parameters.
Transits (circles) and eclipses (diamonds) are shown separately.
Reduced $\chi^2$ values are given in the panel titles. bottom) Combined O$-$C comparison across all four cases.
Left: transits. Right: eclipses.
Case~A (RV-only prediction) is shown with open markers.
Cases B--D show timing residuals from the joint fit.}
\label{fig:oc_residuals}
\end{figure*}

\begin{figure}[h!]
\centering
\includegraphics[width=0.75\textwidth]{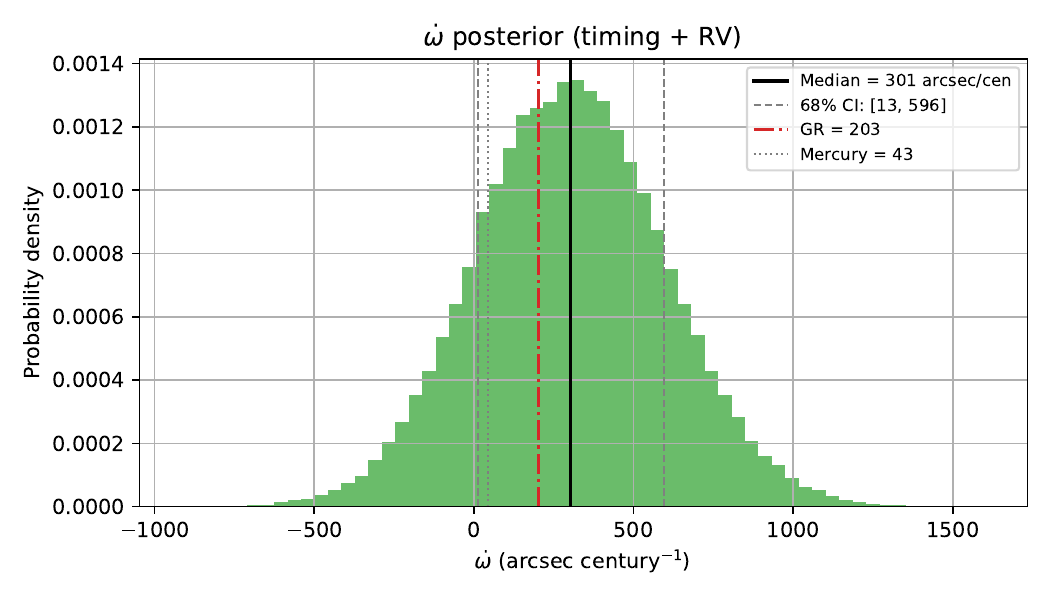}
\caption{Posterior distribution of $\dot\omega$ from Case~D
(RV + all 9 timing events, $\dot\omega$ free).
The median and 68\% credible interval are
$\dot\omega = 302^{+294}_{-289}$\,arcsec\,century$^{-1}$
($1.0\sigma$ from zero).
The GR prediction ($203$\,arcsec\,cen$^{-1}$) and Mercury's
precession rate ($43$\,arcsec\,cen$^{-1}$) are marked.}
\label{fig:omegadot}
\end{figure}

To summarize the above analyses, the combined PRV and transit/eclipse provide excellent fits to the observations with $\chi^2\sim 1$ in all cases. The precision in $T_0$ is greatly improved with the incorporation of the JWST eclipse timing from a few hours (PRV only) to a few minutes. The PRV and eclipse/transit residuals in the four cases are shown in Figures ~\ref{fig:rv_orbits} and \ref{fig:oc_residuals}. The models with and without $\dot{\omega}$ are indistinguishable with nearly identical BIC and AICc values. With its extra free parameter, Case D is slightly disfavored compared to model C, but not ruled out, with $\Delta$AICc=1.08. We attribute the difference between the current analysis and that of \citet{Pearson2022}   to two effects: a poor constraint on $\sqrt{\epsilon}(\cos\,\omega,\, \sin\,\omega)$ with only the single, imprecise Spitzer transit available to them; and  second,  examination of the \citet{Pearson2022} model revealed that the  Spitzer result  was over-weighted  beyond its nominal uncertainty which may also have contributed to the offset.

\begin{table}[h]
\centering
\caption{Model Comparison}
\label{tab:model_comparison}
\begin{tabular}{llccccccc}
\toprule
Case & Data & $N_\mathrm{free}$ & $N_\mathrm{data}$ & RMS & $\ln\mathcal{L}$ & BIC & AICc & $\Delta$AICc \\
     &      &                   &                   & (m\,s$^{-1}$) & & & & \\
\midrule
A & RV only       & 11 & 187 & 5.41 & $-391.5$ & 840.6 & 806.6 & 0.00 \\
B & RV + 7 timing & 11 & 194 & 5.43 & $-393.8$ & 845.6 & 811.1 & 0.00 \\
C & RV + 9 timing & 11 & 196 & 5.47 & $-403.6$ & 865.2 & 830.6 & 0.00 \\
D & RV + 9 timing & 12 & 196 & 5.53 & $-403.2$ & 869.7 & 832.0 & 1.46 \\
\bottomrule
\end{tabular}
\begin{flushleft}
\small
\textbf{Note.}
$\Delta$AICc is computed within groups sharing the same data (C vs.\ D).
Case~A: RV only, $\dot\omega=0$ (fixed).
Case~B: RV + pre-JWST timing (7 events), $\dot\omega=0$.
Case~C: RV + all timing (9 events), $\dot\omega=0$.
Case~D: RV + all timing, $\dot\omega$ free.
All cases include instrument-specific RV jitter terms.
\end{flushleft}
\end{table}

\begin{table}[h]
\centering
\caption{MCMC Posteriors}
\label{tab:FitResults}
\begin{tabular}{lcccc}
\toprule
Parameter & Case~A & Case~B & Case~C & Case~D \\
\midrule
$T_c$ (BJD) & $2458888.012$ & $2458888.077$ & $2458888.078$ & $2458888.078$ \\
             & $\pm9154$\,s  & $\pm132$\,s   & $\pm128$\,s   & $\pm132$\,s   \\[4pt]
$T_\mathrm{peri}$ (BJD)
            & $2458882.356$ & $2458882.350$ & $2458882.361$ & $2458882.361$ \\
             & $\pm1087$\,s  & $\pm342$\,s   & $\pm138$\,s   & $\pm139$\,s   \\[4pt]
$P$ (d)     & $111.43700$   & $111.43703$   & $111.43730$   & $111.43730$   \\
             & $\pm24.4$\,s  & $\pm8.3$\,s   & $\pm1.2$\,s   & $\pm1.2$\,s   \\[4pt]
$e$         & $0.93154^{+0.00058}_{-0.00058}$
            & $0.93153^{+0.00037}_{-0.00037}$
            & $0.93140^{+0.00037}_{-0.00037}$
            & $0.93144^{+0.00038}_{-0.00038}$ \\[4pt]
$\omega_0$ (deg) & $-58.60 \pm 0.23$
                  & $-58.80 \pm 0.11$
                  & $-58.74 \pm 0.11$
                  & $-58.75 \pm 0.11$ \\[4pt]
$\dot\omega$ (arcsec\,cen$^{-1}$)
            & \multicolumn{1}{c}{\ldots}
            & \multicolumn{1}{c}{\ldots}
            & \multicolumn{1}{c}{\ldots}
            & $302^{+294}_{-289}$ \\[4pt]
$K$ (m\,s$^{-1}$)
            & $470.3 \pm 2.4$
            & $470.0 \pm 1.2$
            & $469.7 \pm 1.2$
            & $469.9 \pm 1.3$ \\[4pt]
$\sigma_\mathrm{APF}$ (m\,s$^{-1}$)
            & $4.6 \pm 0.6$
            & $4.7 \pm 0.6$
            & $4.6 \pm 0.6$
            & $4.6 \pm 0.6$ \\[4pt]
$\sigma_\mathrm{HIRES}$ (m\,s$^{-1}$)
            & $3.7 \pm 0.3$
            & $3.7 \pm 0.3$
            & $3.6 \pm 0.3$
            & $3.6 \pm 0.3$ \\[4pt]
$\sigma_\mathrm{HIRES\,pre\,2004}$ (m\,s$^{-1}$)
            & $6.2 \pm 1.0$
            & $6.0 \pm 0.9$
            & $6.3 \pm 0.9$
            & $6.5 \pm 1.0$ \\
\midrule
Timing $\chi^2_\nu$ & \ldots & 0.41 & 2.37 & 2.00 \\
RV $\chi^2_\nu$     & 1.02   & 1.04 & 1.04 & 1.05 \\
Total $\chi^2_\nu$  & 1.02   & 1.01 & 1.11 & 1.09 \\
$N_\mathrm{timing}$ / $N_\mathrm{RV}$ / $N_\mathrm{params}$
                     & 0/187/11 & 7/187/11 & 9/187/11 & 9/187/12 \\
\bottomrule
\end{tabular}
\begin{flushleft}
\small
\textbf{Note.}
Uncertainties are 68\% credible intervals from the MCMC posterior.
$T_c$ is the transit conjunction time at orbit~0 (TESS epoch).
$\omega_0$ is the argument of periastron at the reference epoch
$T_\mathrm{peri} = 2458882.344$\,BJD.
The GR prediction is $\dot\omega_\mathrm{GR} = 203$\,arcsec\,century$^{-1}$.
\end{flushleft}
\end{table}

%% ============================================================
%% FIGURE 1: RV Orbit Plots
%% ============================================================

%% ============================================================
%% FIGURE 2: O-C Residuals
%% ============================================================

%% ============================================================
%% FIGURE 4: omega_dot posterior
%% ============================================================

\section{Long Term changes in Orbital Properties \label{sec:effects}}
The multi-Gyr age of the star suggests that, in the absence of a recent perturbing event, the orbital decay period must be relatively slow for \hdb\ to still be orbiting its host. Equations 9 and 10 of \citet{Hut1981} also suggest that the timescales for changes in period and eccentricity are $>>10^9$ yr. In subsequent sections we discuss possible physical mechanisms for producing changes in  $\omega$: either long-term secular terms ($\dot{\omega}$) due to the presence of a perturbing body, General Relativity, or tidal effects. We also consider short-term timing changes due to transit (eclipse) timing variations due to the presence of a perturbing body.

\subsection{Secular Effects on Periastron}
%\addtocounter{subsection}{-1}%DIFAUXCMD

\begin{figure}[!t]
\centering
\includegraphics[width=0.48\textwidth]{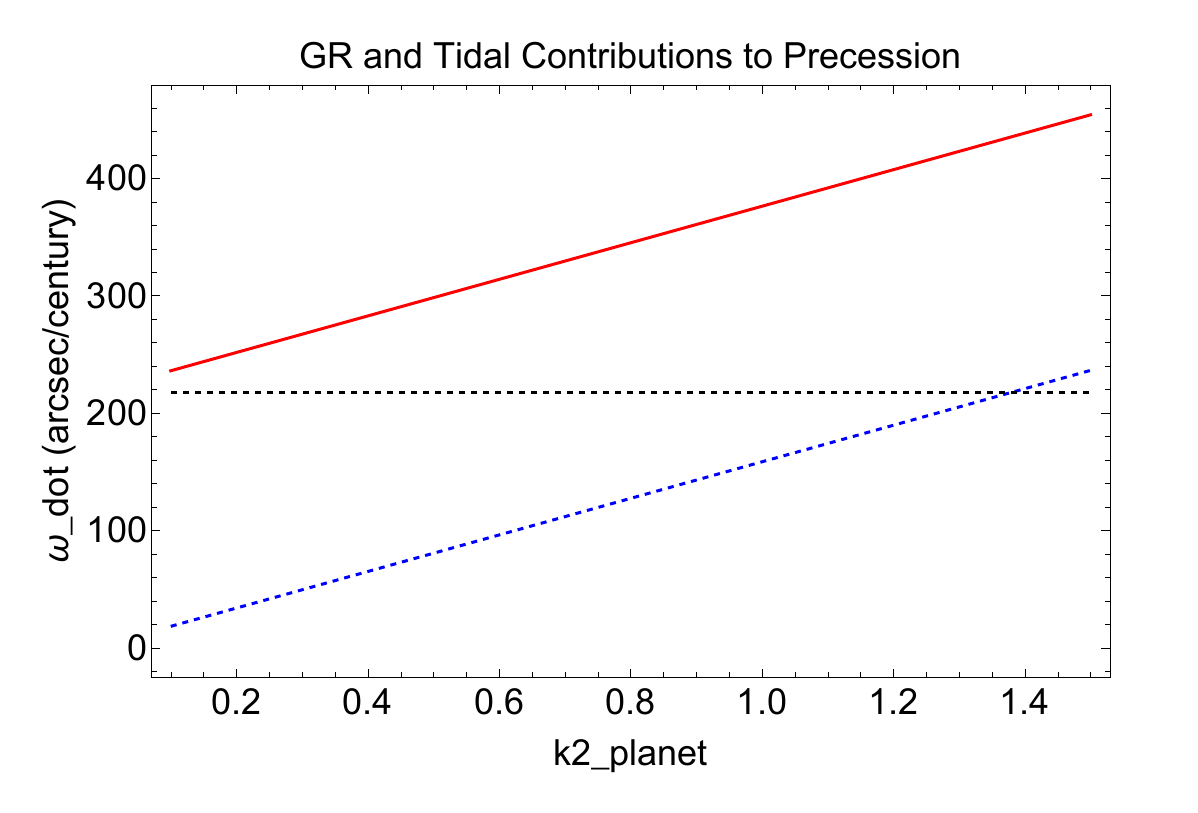}
\caption{Two contributors to $\dot{\omega}$ are GR (black) and the effects of tides (blue). The GR term for \hdb is constant while the tidal effect increases with the $k2_{planet}$ parameter to a maximum value of $\sim$1.5. Together these two effects combine to reach $\sim$400 arcsec century$^{-1}$ (Eqns 2-3).\label{fig:TideGRSums}}
\end{figure}
%\subsubsection{GR effects}
%\addtocounter{subsubsection}{-1}%DIFAUXCMD
\subsubsection{General Relativity Effects}

The first successful validation of General Relativity was its explanation of the 43 arcsec century$^{-1}$ drift in the argument of periastron of the planet Mercury \citep{Einstein1916,Janssen2018}. The effect is described in Eqn~\ref{eqn:GR} \citep{Jordan2008} and is 203 arcsec century$^{-1}$, compared to observed $\dot{\omega}=302^{+294}_{-289}$\,arcsec\,century$^{-1}$.

\begin{equation}
\begin{aligned}
\dot{\omega}_{GR}={} &\frac{3 G M_*}{a c^2(1-e^2)}\frac{2\pi}{P}\\
={}&1395\frac{M_*}{M_\odot} \left(\frac{AU}{a}\right)^{-1} \left(\frac{1}{1-e^2}\right)^{-1} \\
& (P_{day})^{-1} \text{(arcsec \, century}^{-1})\\ %\, \, \text{(deg \, century^{-1})}\\
%={}&203\ \text{(arcsec \, century}^{-1})\\
\label{eqn:GR}
\end{aligned}
\end{equation}

\noindent where symbols have their usual meaning. For the parameters of \hdb, the predicted GR effect, $\dot{\omega}_{GR}=203$ arcsec \, century$^{-1}$, is consistent with the limits of the timing data alone (Column \#4 of Table~\ref{tab:FitResults}) and Figure~\ref{fig:omegadot}.

\subsubsection{Effect of Planetary Tides}

The close approach of \hdb during periastron raises tides on the planet which might  affect the timing of periastron passage as described in Eqn.~\ref{eqn:tide} \citep{Jordan2008}:

\begin{equation}
\begin{aligned}
\dot{\omega}_{tide}={} &1.6 f(e)T(P_{day})^{-1} \left(\frac{k_{2,p}}{0.1}\right)\left(\frac{a}{0.05AU}\right)^{-5} \\
&\left(\frac{R_p}{R_J}\right)^5 
 \frac{M_J}{M_p}\frac{M_*}{M_\odot} \,\, \text{(deg \, century}^{-1})\\
& \text{where} \\
f(e)\equiv{}&(1-e^2)^{-5}\left(1+(3/2)e^2+(1/8)e^4\right) \\
& \text{and} \\
T\equiv{}&1+\left(\frac{R_*}{R_p}\right)^5\left(\frac{M_p}{M_*}\right)^2\left(\frac{k_{2,s}}{k_{2,p}}\right)
\label{eqn:tide}
\end{aligned}
\end{equation}

\noindent where symbols have their usual meaning with $k_{2,p}$ and $k_{2,s}$ referring to second-order shape deformation constants for the planet and star respectively. $k_{2,s}$ has a typical value of 0.01 while $k_{2,p}$ can vary between 0.1 to 1.5 \citep{Jordan2008}. The contribution of tidal effects to $\dot{\omega}$ is driven by the value of $k_{2,p}$ as shown in Figure~\ref{fig:TideGRSums}. At the theoretical maximum value of $k_{2,p}\sim 1.5$, the tidal $\dot{\omega}\sim200$ arcsec century$^{-1}$ and the tidal to GR ratio is roughly equal to 1 for a combined $\dot{\omega}\sim$ 400 arcsec century$^{-1}$. Subtracting the GR term from the observed posteriors yields a limit to $\dot{\omega}_{\rm{tidal}}<99^{+282}_{-289}$.

\section{Limits on the Presence of a Second Planet\label{sec:second_planet}}

\begin{figure*}[!t]
\centering
\includegraphics[width=0.8\textwidth]{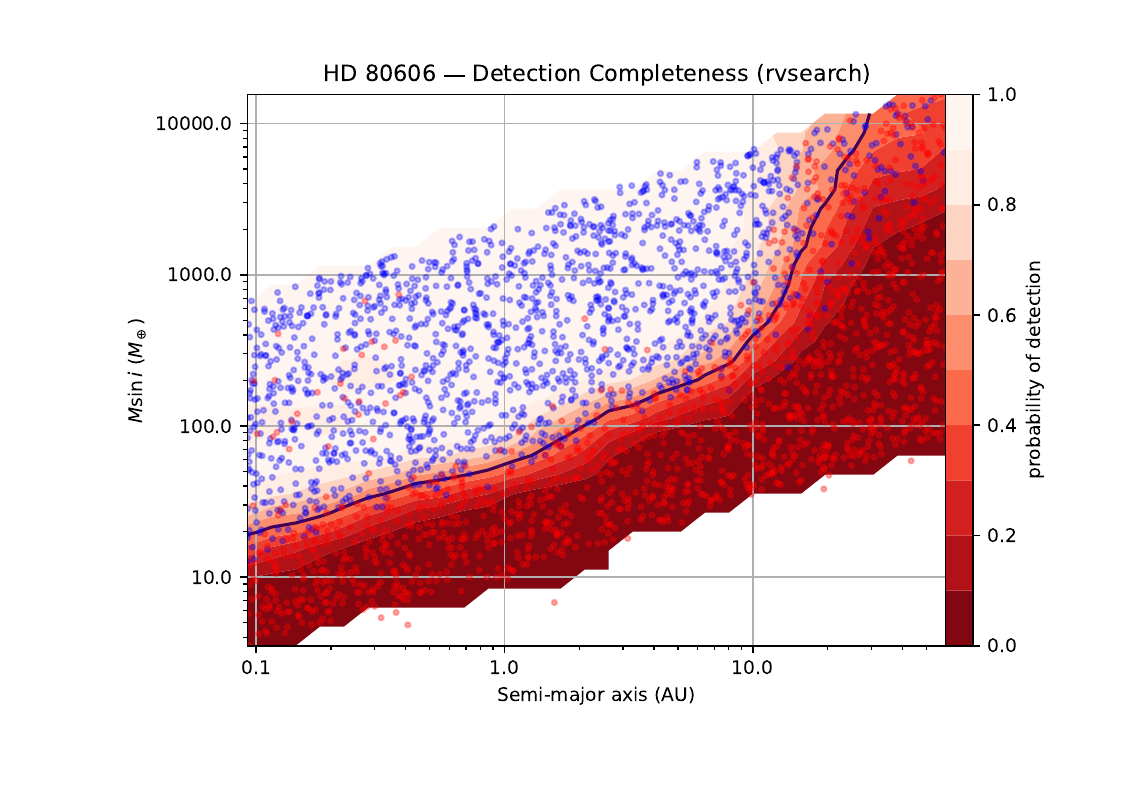}
\caption{An injection-recovery analysis of the PRV  after subtracting the signature of \hdb  shows mass limits on a second planet in the system as a function of orbital semi-major axis.\label{fig:injection}}
\end{figure*}

The high eccentricity of \hdb might originally have been  caused by the presence of a second planet in the system. Such a planet might have long ago been ejected \citep{Carrera2019} or might still be present. In this section we address how such a planet, \hdc, might manifest itself today.  A sufficiently massive planet would produce a detectable PRV signal  once the signature of \hdb has been removed. Figure~\ref{fig:injection} shows the results of an injection/recovery analysis of the PRV residuals, setting a limit on a planet at 2 au of $\sim$100 \mearth. A second planet could also manifest itself more subtly by changing the timing of \hdb's eclipses and transits, either producing a long term drift in $\omega$ or as  sporadic changes due to their mutual interactions, i.e. as Transit or Eclipse Timing Variations (ETV/TTVs). 

Following \citet{Jordan2008} we estimate the long-term secular effect of a second planet on the orbit of \hdb:

\begin{equation}
\begin{aligned}
 \dot{\omega}_{pertrub}={}&\frac{3\,M_2\, a^3}{4\, M_*\,a_2^3}\frac{2\pi}{P}\\
 ={}&29.6 P_d^{-1}\left(\frac{a}{a_2}\right)^3\,\frac{M_\odot}{M_*}\,\frac{M_2}{M_\oplus}\\
 & \text{(deg century}^{-1})
\end{aligned}
\end{equation}
\vspace{0.1 in}

\noindent where $M_2$ and $a_2$ are the mass and semi-major axis of the perturbing planet. Inserting the properties of \hdb into the above gives  $\dot{\omega}=93\,  a_2^{-3}\, M_{2,\oplus}$ arcsec century$^{-1}$. For example, a 100 \mearth planet at 2 au would produce $\dot{\omega}=1200$   arcsec century$^{-1}$ which is ruled out by our limits (Figure~\ref{fig:omegadot}).

 More generally, the presence of a perturbing planet would produce non-secular changes in eclipse and transit timing, i.e. ETVs and TTVs, as the planets interact \citep{Agol2005,Holman2005,Nesvorny2008,Lithwick2012,Borkovits2016}. We investigate this region using the n-body software package \texttt{rebound} (v.4.4.6) \citep{Rein2012,Rein2015} and its extension package \texttt{reboundX} (v.4.4.1) \citep{Tamayo2020}.  \texttt{rebound} is the industry standard for high-precision numerical integration of orbits due to its high-order adaptive-step method \texttt{IAS15} that accurately integrates highly eccentric orbits while keeping the wall-time of the simulation low.  For \hdb, the eccentricity $(e_b \approx 0.93)$ is very high, which places its periastron distance ($q_b \approx 0.032$ au) very close to the host star.  Thus, we must include the effects of general relativistic (GR) precession in our integration, which is handled by \texttt{reboundX} through its \texttt{gr} module that correctly models the GR precession and the mean motion for objects near the central body (i.e., HD80606 + \hdb).  Note this is not equivalent to the full potential, but these effects are small for the putative planet and the binary companion.

This software is used to evaluate the 3-body system (host star + planet b + planet c) for possible TTVs and ETVs of planet b.  Using \texttt{rebound} and \texttt{reboundX}, we perform an event search that minimizes the sky-plane separation (i.e., identify conjunctions where the impact parameter $b\lesssim 1$) until 60 transits/eclipses are identified.  This timescale is much less than the estimated period of the binary companion, which justifies neglecting it altogether from this analysis.  We perform a linear fit to the event times to determine the TTVs \citep{Agol2005,Holman2005}, while a quadratic fit is used for the eclipses.  The higher-order fitting is necessary because of the shorter duration of the eclipse events which makes the timing measurement more sensitive to small variations in the sky-plane separation.  Note that we did try a linear fit for the eclipses initially, but this resulted a floor in the ETV times of 30 minutes due to sporadic or missed events.  Also, this method produced a spuriously high variation in the first event, which we discarded and only fit the remaining events.  Our method does include the light-travel time effect typically used for eclipsing binaries \citep{Borkovits2016} for completeness; however, this contribution is negligible ($t_{\rm LTTE} \ll 1\ {\rm s}$) within the 3-body system.

\begin{figure*}[!t]
\centering
\includegraphics[width=0.98\textwidth]{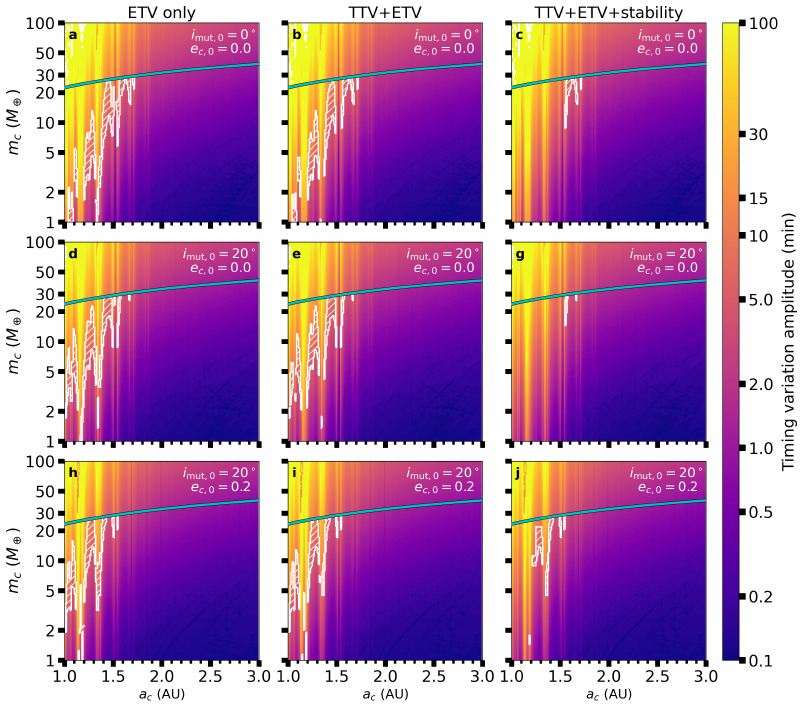}
\caption{Timing variations determined using \texttt{rebound}+\texttt{rebound}x (3-body system) considering the eclipse timing variations (ETVs; panels a, d, \& h), and transit timing variations (TTVs; panels b, c, e, g, i, \& j) from a second planet $c$ with a mass $m_c$ and initial semi-major axis $a_c$. The cyan line represents the upper limit set by the RV measurements.  The (white) hatched region in the ETV only plots mark which initial parameters are consistent with the ETV observations, while the hatched region in the TTV+ETV column combine the constraints.  The hatched region in the TTV+ETV+stability column show which initial conditions have $>50\%$ stability over a 3 Myr timescale.\label{fig:TTV}}
\end{figure*}

Figure \ref{fig:TTV} shows the results of this analysis using a color-coded (by the timing variation) map of the parameter space.  In these simulations, we sample the putative mass $m_c$ in logarithmic steps of $\log_{10}(m_c/M_\oplus)$ from 0 to 2.0 in steps of 0.01, while sampling the semi-major axis $a_c$ from 1 to 3 in steps of 0.01 au.  Since the binary companion was not included, we performed simulations assuming the two planets are coplanar or misaligned by $20^\circ$.  We also search for TTV/ETV signatures from a planet c on an initially circular and eccentric ($e_c=0.2$) orbit.  We overlay a hatched region (in white) to identify those conditions that could potentially match the observed bounds on the TTVs/ETVs.  

\begin{figure*}[!t]
\centering
\includegraphics[width=\textwidth]{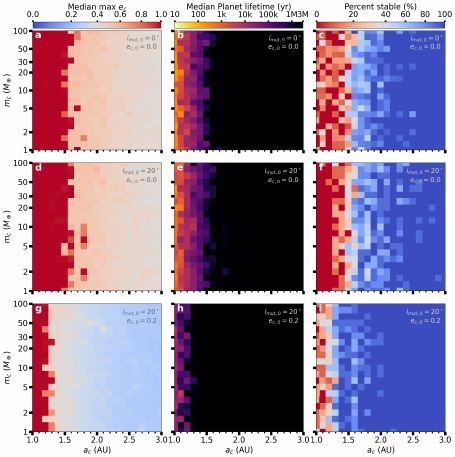}
\caption{Coarse stability diagnostics using \texttt{rebound}+\texttt{rebound}x (4-body system), where each cell summarizes 10 simulations with the relative initial phase between the planets is randomized. Panels a, d, \& g show the median maximum eccentricity reached by planet $c$ during each integration.  Panels b, e, \& h illustrate the median lifetime of the system, while panels b, e, \& h show the fraction of integrations that remain stable over a 3 Myr timescale. \label{fig:stability}}
\end{figure*}

We also simulate the 4-body system (host star + 2 planets + binary companion) up to 3 Myr assuming that the binary is on a circular orbit with a separation of $1200$ au.  Since we do not know the relative orbital phase between the planets, we reduce the grid resolution so that we evaluate steps of 0.1 in the logarithmic mass and 0.1 au steps for $a_c$.  Within this lower resolution, we simulate 10 different initial mean anomalies for planet c to ensure that we are not biased towards any particular (favorable) starting condition \citep{Quarles2020}.  Note that the two planets begin with their arguments of pericenter aligned, $\omega_b = \omega_c = 0$.  Figure \ref{fig:stability} shows the maximum eccentricity attained by planet c, the median lifetime (up to the 3 Myr simulation time), and the percentage of initial conditions that survived the whole simulation time (i.e., provisionally stable). We find that the eccentricity of planet c ($e_c$) is forced by the inner planet b \citep{Andrade-Ines2016,Quarles2018a,Quarles2026} and the maximum eccentricity is $e_c {\sim}0.5$, where this is favorable for inducing a possible TTV/ETV signature.  However, most of the simulations in Figs. \ref{fig:stability}c and \ref{fig:stability}f that allow for $\gtrsim50\%$ stability require a semi-major axis $a_c>2$ au, which does not match the TTV/ETV constraints.  If we assume that planet c is mutually inclined relative to planet b, and planet c begins with a moderate $e_c=0.2$ eccentricity, a region from $1.5\ {\rm au} < a_c < 2\ {\rm au}$ is provisionally stable in Fig. \ref{fig:stability}i and can match the TTV/ETV constraints.  This places limits on the mass of the putative planet c as $3\ M_\oplus\lesssim M_c \lesssim 25 M_\oplus$, albeit non-uniformly relative to the vertical teeth caused by strong mean motion resonances \citep{Nesvorny2008,Lithwick2012} seen in Fig. \ref{fig:TTV}.

\subsection{Prospects for Future Observations}

Although JWST eclipse or transit measurements have  a timing uncertainty of a few 10s of seconds, the net timing accuracy needed  for comparison with ETv/TTVs  or with models with secular drifts ($\dot{\omega}>0$) is dominated by the uncertainty in $T_0$ and $P$, 130 sec and 3 sec, respectively (Table~\ref{tab:FitResults}): $\sigma=\sqrt{\sigma_{JWST}^2+\sigma_{T_0}^2+(n\sigma_P)^2}$.  \citet{Blanchet2019} point out that observations of an eclipse followed by a transit \textit{in the same orbit} would provide a robust method looking for long term drifts independent of $T_0$ and $P$. The change in this timing interval depends primarily on $\epsilon$ and $\omega$ (Equation~\ref{eqn:timing}) with $P$ entering only as a scaling factor. We define $\Delta$ET as the interval between an eclipse and the subsequent transit about 5.8 days later:

\begin{equation}
\begin{aligned}
\Delta ET(n)={}&\int_{0}^{\pi/2-(\omega_0+n\times \dot{\omega})}f(x) dx-
\int_{0}^{-\pi/2-(\omega_0+n\times \dot{\omega})}f(x) dx% \\
%\Delta TE(n)-\Delta TE(0)={}&n\, \frac{2 \epsilon \left(1-\epsilon^2\right)^{3/2} %\dot{\omega} P \sin (\omega_0)}{\pi  \left(\epsilon^2 \sin ^2(\omega_0)-1\right)^2} 
\label{eqn:fitfit}
\end{aligned}
\end{equation}
\noindent where $f(x)$ incorporates the terms given in Equation~\ref{eqn:timing} and $n$ is the orbit number. Using the fundamental theorem of calculus and some simplifications, we show  that the time between an eclipse and the succeeding transit changes with increasing orbit number:

\begin{equation}
\begin{aligned}
\Delta ET(n)-\Delta ET(0)={}&n\, \frac{2 \epsilon \left(1-\epsilon^2\right)^{3/2} \dot{\omega} P \sin (\omega_0)}{\pi  \left(\epsilon^2 \sin ^2(\omega_0)-1\right)^2} \label{eqn:fitfit2}
\end{aligned}
\end{equation}

 \noindent where we have assumed $\omega\approx\omega_0$ in the $sin$ term.  $\Delta$ET can be measured with JWST's full precision with  only small uncertainties introduced by the well determined orbital parameters, $P$ and $\epsilon$. Measuring  $\Delta ET$ at intervals of, say, $n=0,5,10,20$ periods, i.e. over about 6 years, would yield a  measurement of $\dot{\omega}  = 400\pm$50 arcsec century$^{-1}$ due to a combination of GR and (possibly) tidal effects (Figure~\ref{fig:TideGRSums}). While the combined measurements would require approximately 24 hours per transit$+$eclipse pair, the result would yield multiple atmospheric spectra to look for variability and, potentially, a first measurement of tidal effects in a highly eccentric giant planet.

\begin{figure*}[!t]
\centering
\includegraphics[width=\textwidth]{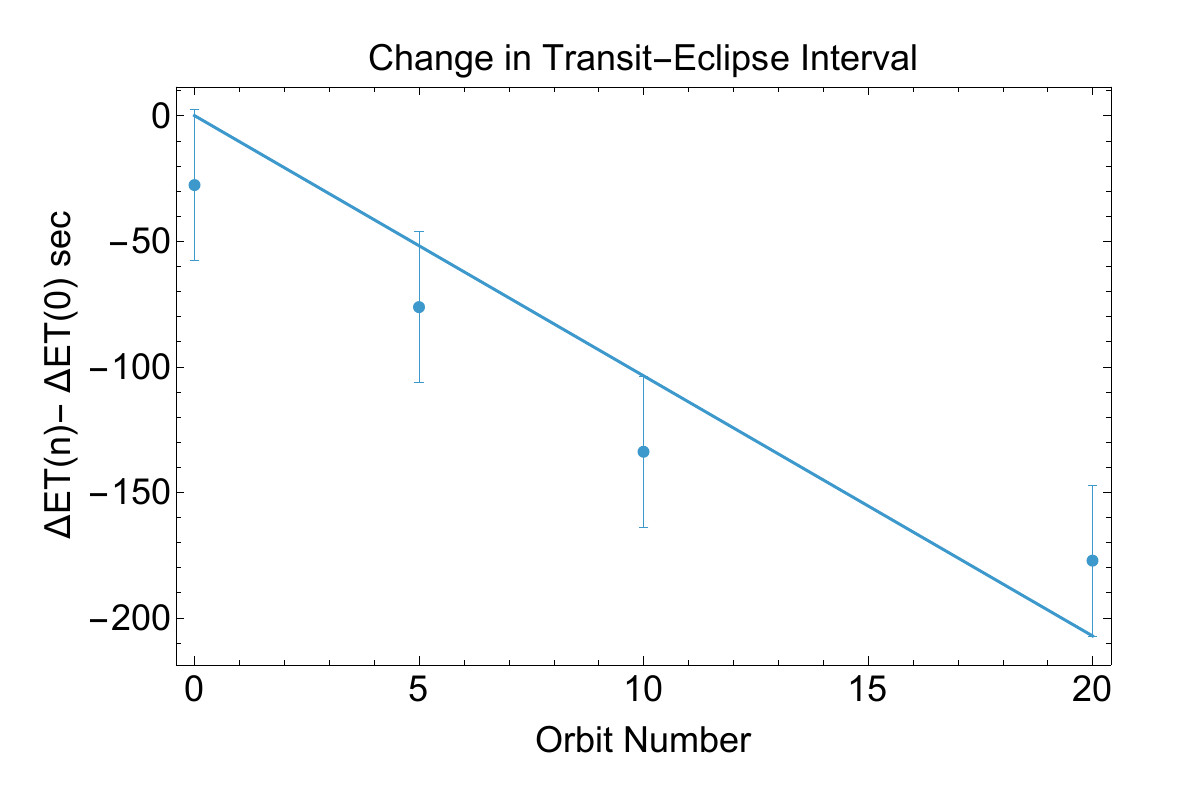}
\caption{Four simulated JWST measurements with a combined uncertainty of 30 sec of the evolving interval between an  eclipse and its immediately following transit ($\Delta ET$; Equation~\ref{eqn:fitfit2}) over a span of 20 orbits (approximately 6 years). Each measurement is taken to have a combined eclipse+transit uncertainty of 30 sec and the final result would yield a 10\% measurement of $\dot{\omega}=$ 400 arcsec century$^{-1}$, enough to make a 5$\sigma$ detection of a tidal effect after correcting for the GR term. \label{fig:FitOmega}}
\end{figure*}

\section{Conclusions \label{sec:conclusion}}

We have combined archival PRV, transit and eclipse timing data with a new eclipse observation using JWST/MIRI and a new transit observation from CHEOPS to refine the orbital parameters of the highly eccentric planet \hdb. We find that a model with constant period, eccentricity and $\omega$ provides an excellent fit to the data.  A  model with variable $\omega(t)=\omega_0+t \dot{\omega}$ also fits the data but places only a limit on $\dot{\omega}$.  We explored the effect of the presence of a previously undetected planet looking at both residuals in the PRV data  and transit/eclipse timing variations. The PRV data set a firm limit on the properties of such a planet, M$\lesssim$100 \mearth\ orbiting \hd\ in the $>2$ au range. The lack of measurable TTV/ETVs put more stringent constraints but the cadence of these observations is too small to be conclusive. Stability tests extending over 3 Myr and current RV limits favor a low mass and larger SMA of any perturber. Finally, we note that these parameters presented differ from timing predictions resented in \citet{Pearson2022} due to the availability of only the  single eclipse observation (Spitzer in 2009) which was over-weighted in that analysis. A new table of timing of eclipses and transits over the next 10 years is presented in $\S$~Appendix C (Table~\ref{tab:ephemeris}).

\facilities{JWST(MIRI/LRS), JWST(NIRSpec), CHEOPS, NASA Exoplanet Archive, Keck(HIRES), Lick(APF), Keck Observatory Archive (KOA)}

\software{
\texttt{astropy} 
\citep{astropy2013, astropy2018, astropy2022}, 
\texttt{emcee} \citep{FM2013},
\texttt{Eureka!+batman} \citep{Kreidberg2015},
\texttt{EXOFAST} \citep{Eastman2013},
\texttt{RADVEL} \citep{Fulton2018},
\texttt{Eureka!+batman} \citep{Kreidberg2015}
}

\noindent  The MIRI data associated with PID \# 2008 are available at: 
\dataset[https://doi.org/10.17909/91vc-1r62]{https://doi.org/10.17909/91vc-1r62
}\\
The PRV data (Table~\ref{tab:PRV_data}) are available at:
\dataset[https://doi.org/10.5281/zenodo.20096135]{https://doi.org/10.5281/zenodo.20096135
}

\section{Acknowledgments}

C.A.B. thanks Victoria Meadows for the hospitality of the University of Washington during a short sabbatical stay and acknowledges useful early discussions about \hd with Eric Agol. Some of the research described in this publication was carried out in part at the Jet Propulsion Laboratory, California Institute of Technology, under a contract with the National Aeronautics and Space Administration (80NM0018D0004). The JWST GO program \#2008 is supported  by grant JWST-GO-02008. This research has made use of the NASA Exoplanet Archive and ExoFOP, which is operated by the California Institute of Technology, under contract with the National Aeronautics and Space Administration under the Exoplanet Exploration Program.  MR acknowledges support from Heising-Simons Foundation Grants \#2021-2802 and \#2023-4478, as well as the NASA Exoplanets Research Program NNH23ZDA001N-XRP (grant \#80NSSC24K0153).

Some of the data presented herein were obtained at the W. M. Keck Observatory, which is operated as a scientific partnership among the California Institute of Technology, the University of California and the National Aeronautics and Space Administration. The Observatory was made possible by the generous financial support of the W. M. Keck Foundation.  The authors wish to recognize and acknowledge the very significant cultural role and reverence that the summit of Maunakea has always had within the indigenous Hawaiian community.  We are most fortunate to have the opportunity to conduct observations from this mountain.

\clearpage

\appendix
%\section{CHEOPS Transit}
%{\color{red} The derived parameters from the EXOFAST fit to the CHEOPS light curve. 
%\textbf{Should we include?}}
%\input{exofast.median.e.f.tex}
%\clearpage
\section{Model Fitting Corner Plots}

Corner plots for the four model cases discussed in the main body of the paper.
%% ============================================================
%% FIGURE 5: Corner Plots
%% ============================================================
\begin{figure*}[h]
\centering
\includegraphics[width=0.45\textwidth]{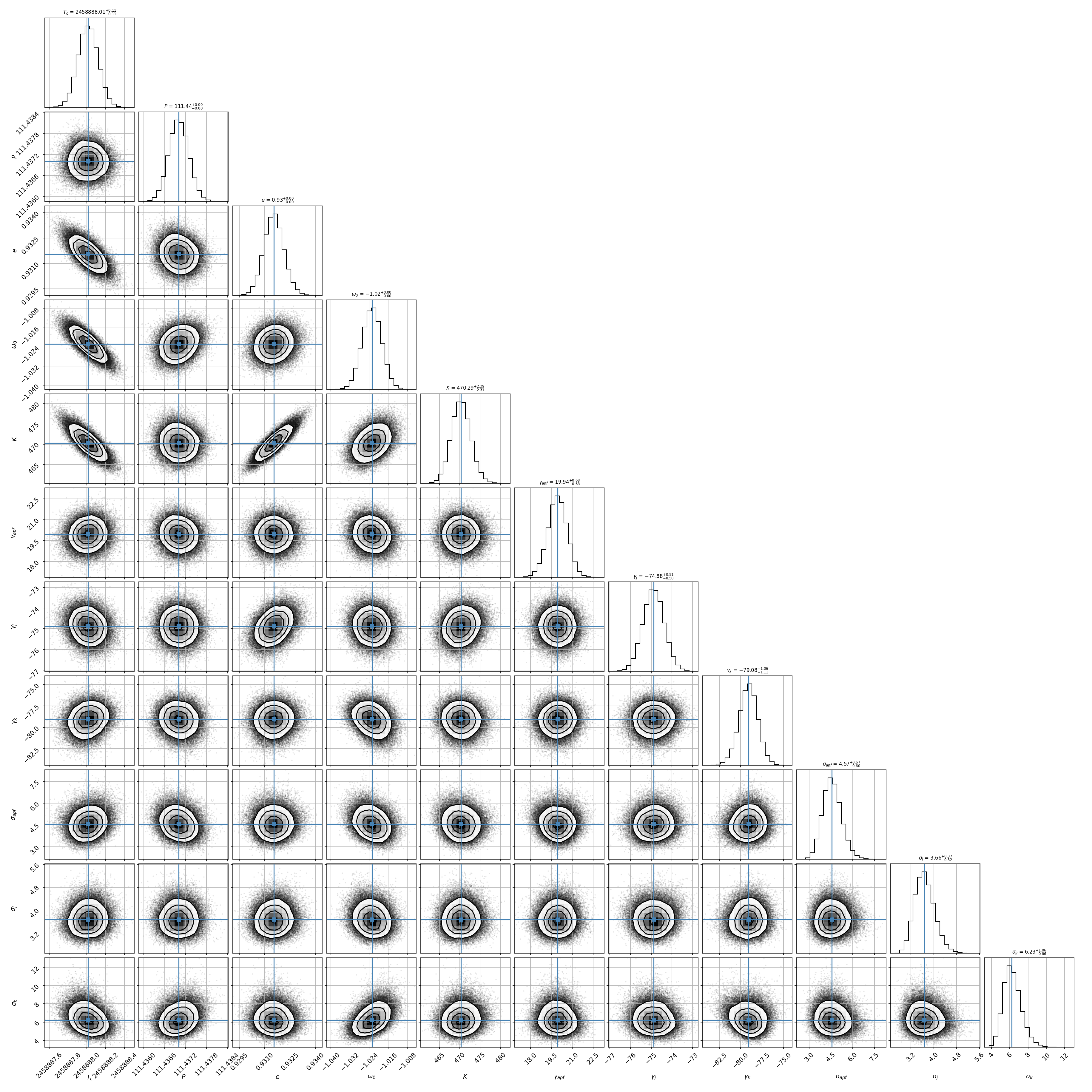}
\includegraphics[width=0.45\textwidth]{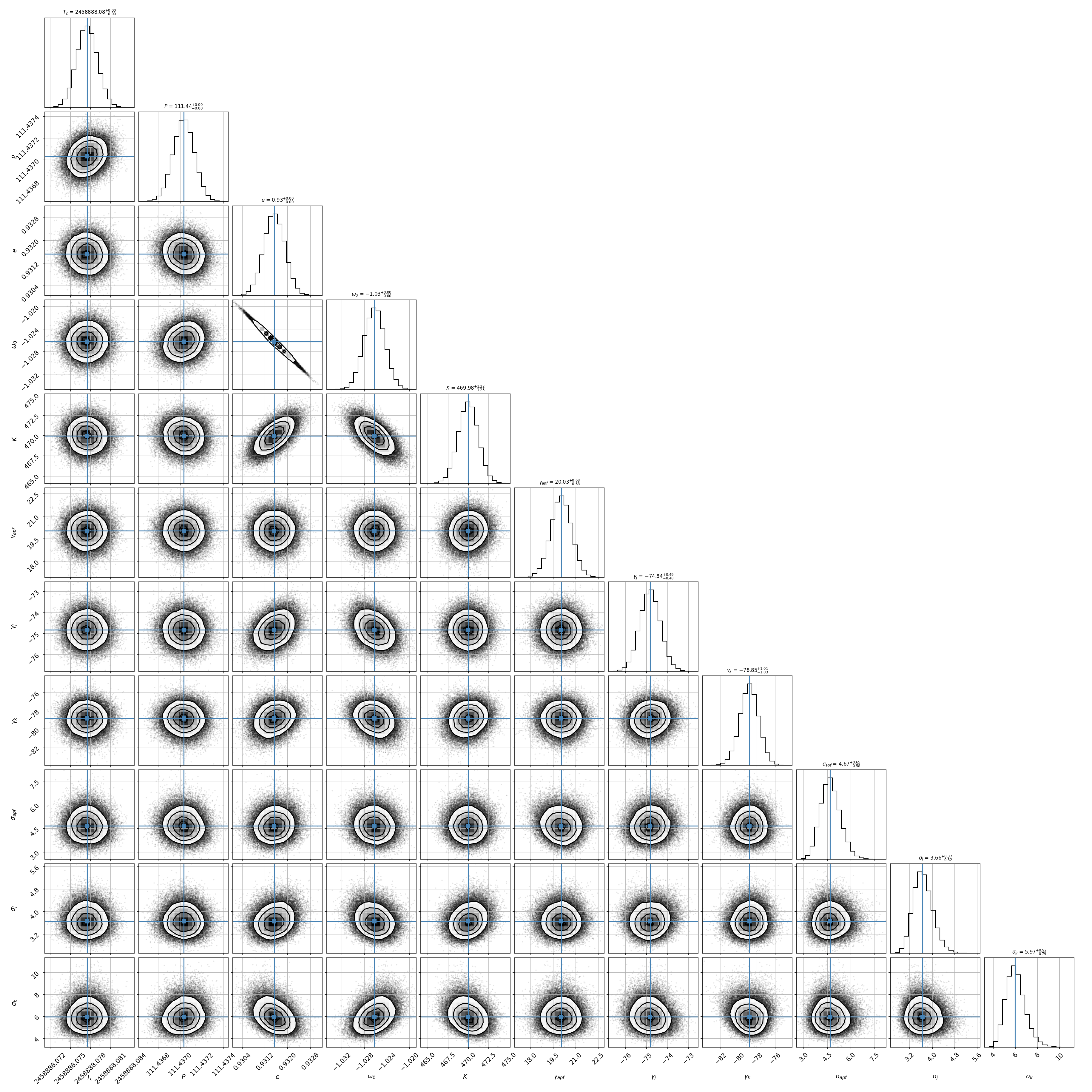}\\[6pt]
\includegraphics[width=0.45\textwidth]{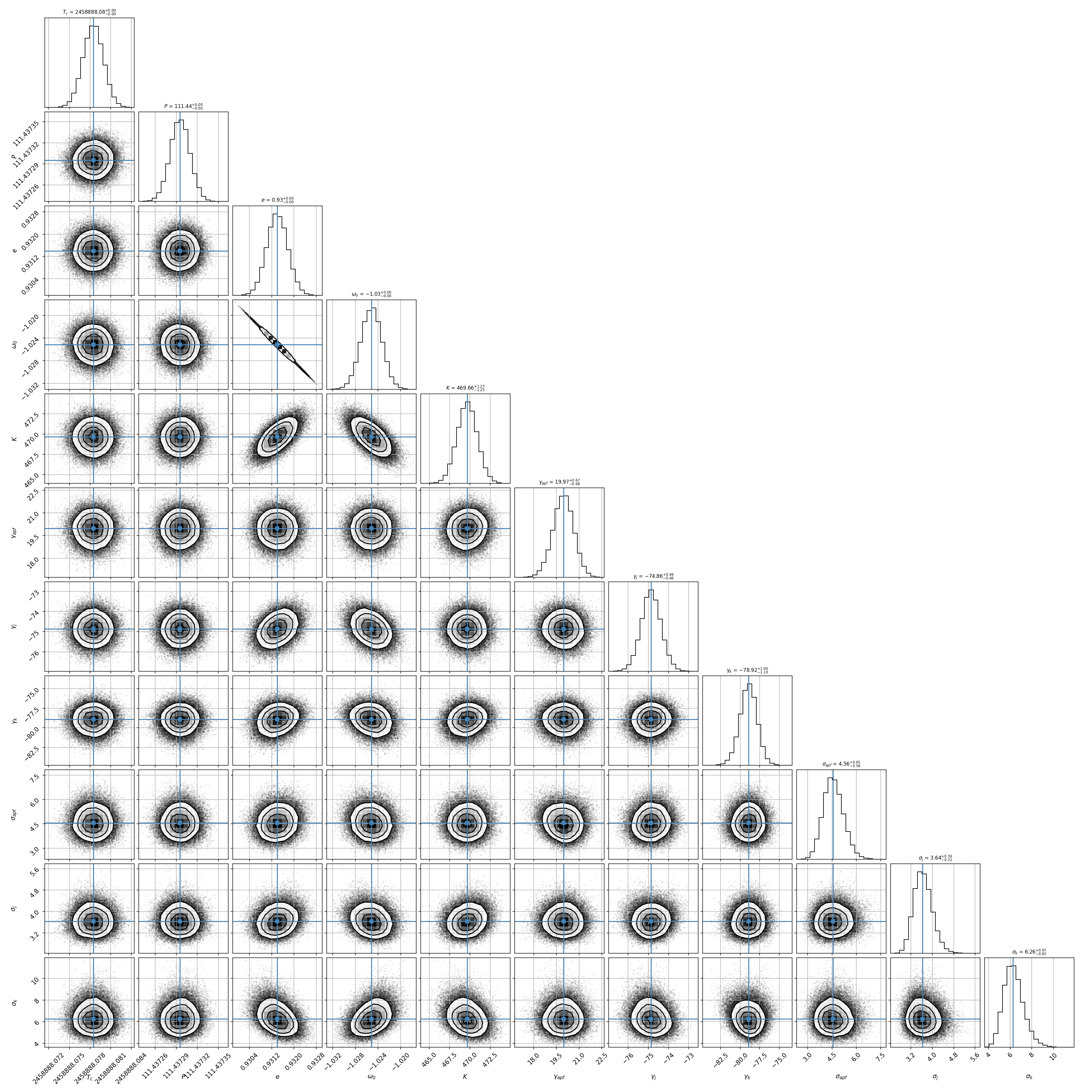}
\includegraphics[width=0.45\textwidth]{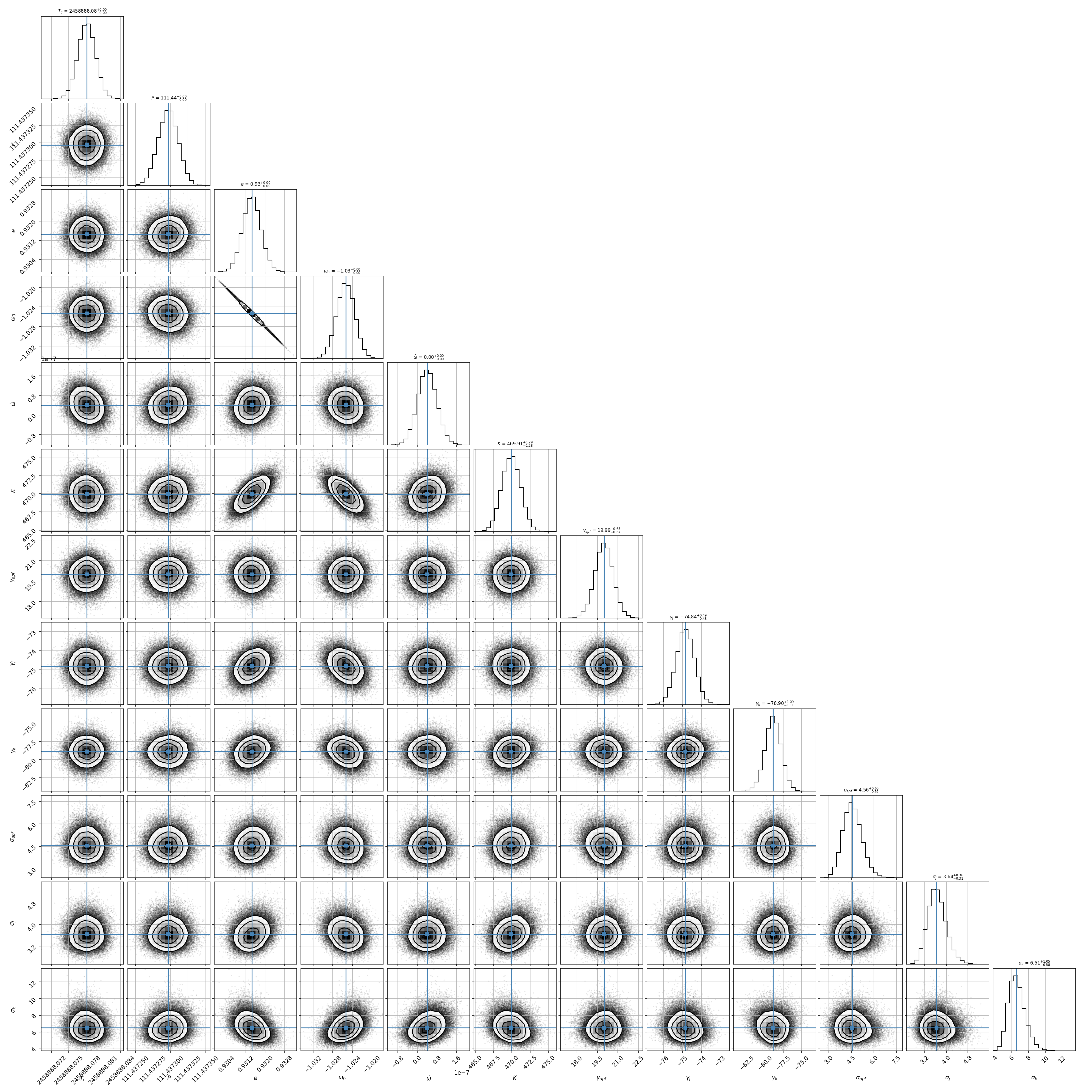}
\caption{Corner plots showing the posterior distributions for
all fitted parameters in each case.
Case~D (bottom right) includes $\dot\omega$ as an additional free parameter.}
\label{fig:corners}
\end{figure*}

%\clearpage
\clearpage
\section{Predicted Times of Periastron Passage, Eclipse and Transit mid-points \label{sec:PredcitedTimes}}

Table~\ref{tab:ephemeris} uses model parameters from Case C ($\dot{\omega}=0$) from Table~\ref{tab:FitResults}. The event uncertainty on the $nth$ orbit is dominated by the uncertainties in T$_0$ and period, P, $\sigma(T_n)=\sqrt{\sigma (T_0)^2+(n \sigma (P))^2}$. Uncertainties due to $\omega$ and eccentricity are negligible in comparison. 
This table is an updated version of the table in \citet{Pearson2022} with improved knowledge due to the precise JWST eclipses and an appropriately weighted Spitzer transit.

\begin{deluxetable}{llccc}[b!]
\centering
\tabletypesize{\scriptsize}
\tablecaption{Predicted Event Times for \hdb\label{tab:ephemeris}}
\tablehead{
\colhead{Orbit} & \colhead{Periapsis Date} & \colhead{T(Periapsis) BJD$_{TBD}$}& \colhead{T(Eclipse) BJD$_{TBD}$} & \colhead{T(transit) BJD$_{TBD}$ }}
\startdata
%Orbit& Periapsis & TPeri (BJD_TBD) &Emid (BJD_TBD)& Tmid \\
0 &2020-2-2 20:39:50 &2458882.3610(128.s)& 2458882.2289& 2458888.0768\\
1 &2020-5-24 07:09:24 &2458993.7982(128.s)& 2458993.6661& 2458999.5140\\
2 &2020-9-12 17:38:59 &2459105.2354(128.2s)& 2459105.1033& 2459110.9512\\
3 &2021-1-2 04:08:33 &2459216.6726(128.3s)& 2459216.5405& 2459222.3884\\
4 &2021-4-23 14:38:07 &2459328.1098(128.6s)& 2459327.9777& 2459333.8256\\
5 &2021-8-13 01:07:41 &2459439.5470(128.9s)& 2459439.4149& 2459445.2628\\
6 &2021-12-2 11:37:15 &2459550.9842(129.3s)& 2459550.8521& 2459556.7000\\
7 &2022-3-23 22:06:49 &2459662.4214(129.8s)& 2459662.2893& 2459668.1372\\
8 &2022-7-13 08:36:23 &2459773.8586(130.4s)& 2459773.7265& 2459779.5744\\
9 &2022-11-1 19:05:57 &2459885.2958(131.s)& 2459885.1637& 2459891.0116\\
10 &2023-2-21 05:35:31 &2459996.7330(131.7s)& 2459996.6009& 2460002.4488\\
11 &2023-6-12 16:05:05 &2460108.1702(132.5s)& 2460108.0381& 2460113.8860\\
12 &2023-10-2 02:34:39 &2460219.6074(133.3s)& 2460219.4753& 2460225.3232\\
13 &2024-1-21 13:04:13 &2460331.0446(134.2s)& 2460330.9125& 2460336.7604\\
14 &2024-5-11 23:33:48 &2460442.4818(135.2s)& 2460442.3497& 2460448.1976\\
15 &2024-8-31 10:03:22 &2460553.9190(136.2s)& 2460553.7869& 2460559.6348\\
16 &2024-12-20 20:32:56 &2460665.3562(137.3s)& 2460665.2241& 2460671.0720\\
17 &2025-4-11 07:02:30 &2460776.7934(138.4s)& 2460776.6613& 2460782.5092\\
18 &2025-7-31 17:32:04 &2460888.2306(139.6s)& 2460888.0985& 2460893.9464\\
19 &2025-11-20 04:01:38 &2460999.6678(140.9s)& 2460999.5357& 2461005.3836\\
20 &2026-3-11 14:31:12 &2461111.1050(142.2s)& 2461110.9729& 2461116.8208\\
21 &2026-7-1 01:00:46 &2461222.5422(143.6s)& 2461222.4101& 2461228.2580\\
22 &2026-10-20 11:30:20 &2461333.9794(145.s)& 2461333.8473& 2461339.6952\\
23 &2027-2-8 21:59:54 &2461445.4166(146.5s)& 2461445.2845& 2461451.1324\\
24 &2027-5-31 08:29:28 &2461556.8538(148.1s)& 2461556.7217& 2461562.5696\\
25 &2027-9-19 18:59:02 &2461668.2910(149.6s)& 2461668.1589& 2461674.0068\\
26 &2028-1-9 05:28:36 &2461779.7282(151.3s)& 2461779.5961& 2461785.4440\\
27 &2028-4-29 15:58:11 &2461891.1654(152.9s)& 2461891.0333& 2461896.8812\\
28 &2028-8-19 02:27:45 &2462002.6026(154.7s)& 2462002.4705& 2462008.3184\\
29 &2028-12-8 12:57:19 &2462114.0398(156.4s)& 2462113.9077& 2462119.7556\\
30 &2029-3-29 23:26:53 &2462225.4770(158.2s)& 2462225.3449& 2462231.1928\\
31 &2029-7-19 09:56:27 &2462336.9142(160.1s)& 2462336.7821& 2462342.6300\\
32 &2029-11-7 20:26:01 &2462448.3514(161.9s)& 2462448.2193& 2462454.0672\\
33 &2030-2-27 06:55:35 &2462559.7886(163.9s)& 2462559.6565& 2462565.5044\\
34 &2030-6-18 17:25:09 &2462671.2258(165.8s)& 2462671.0937& 2462676.9416\\
35 &2030-10-8 03:54:43 &2462782.6630(167.8s)& 2462782.5309& 2462788.3788\\ \hline
\enddata
\tablecomments{Periastron times, $T_{peri}$, are predicted on the basis of the $T_0$, Period and orbit number. Eclipse and Transit times are predicted from $T_{peri}$ using Equation~\ref{eqn:timing} with orbit parameters taken from (Case C in Table~\ref{tab:FitResults}). The uncertainties in $T_{peri}$, $T_{eclipse}$, $T_{transit}$ are dominated by the 130 sec uncertainty in $T_{peri}$ with a slowly growing contribution from the uncertainty in the period. }
\end{deluxetable}

\end{document}